\documentclass{aa}

\usepackage{graphicx}
\usepackage{txfonts}
\usepackage{siunitx}

\usepackage{lineno}
\usepackage{xcolor}

\usepackage[normalem]{ulem}
\usepackage{amsmath}
\usepackage{hyperref}
\usepackage{tablefootnote}
\usepackage{float}

\hypersetup{
  colorlinks=true,
  linkcolor=blue,
  urlcolor=blue,
  citecolor=blue
  }

\DeclareSIUnit\erg{erg}
\DeclareSIUnit\dyne{dyn}
\DeclareSIUnit\dn{DN}
\DeclareSIUnit\maxwell{Mx}

\DeclareSIUnit\angstrom{\text {Å}}
\DeclareUnicodeCharacter{03B2}{$\beta$}

\DeclareSIUnit\gauss{G}

\begin{document}

\newcommand{\software}{\textit}
   \title{Small-scale impulsive extreme-UV emission enhancements along network loops}

   \author{A. Dolliou \inst{\ref{aff:mps}} 
    \and H. Peter
    \inst{\ref{aff:mps},\ref{aff:kis}}
    \and S. Mandal
    \inst{\ref{aff:mps}}
    \and L. P. Chitta
    \inst{\ref{aff:mps}}
    \and L. Teriaca 
    \inst{\ref{aff:mps}}
    \and Y. Chen
    \inst{\ref{aff:mps}}  
    \and D. Calchetti
    \inst{\ref{aff:mps}} 
          }

    \institute{%
        \label{aff:mps}{Max Planck Institute for Solar System Research, Justus-von-Liebig-Weg 3, 37077 Göttingen, Germany}
    \\ \email{dolliou@mps.mpg.de}
    \and    
        \label{aff:kis}{Institute for Solar Physics (KIS), Georges-K{\"o}hler-Allee 401A, 79110 Freiburg, Germany}
    }

   \date{Accepted on; published on}

 
  \abstract
  {Network loops are a common feature in the quiet Sun. The physical processes sustaining their energy budget are still discussed.} 
   {We relied on a multi-instrumental (Solar Orbiter/EUI, Solar Orbiter/PHI, and IRIS) observation of a six-hour quiet-Sun region to measure the dynamics and the possible magnetic drivers of impulsive Extreme Ultraviolet (EUV) emission enhancements along network loops.}
   {We report the detection of small-scale impulsive EUV emission enhancements with EUI/HRIEUV in three network loops. We selected four EUV emission enhancements to measure their plane-of-sky velocities in HRIEUV, their Doppler velocities in the \ion{Si}{iv} line ($\log{T} = 4.8$) with IRIS, and their possible relation to small-scale flux emergence and fluctuation in one of the loop footpoints. }
   {The plane-of-sky velocities of the four EUV emission enhancements have a component that seems to appear almost instantaneously along the loop ($\geq$ \SI{220}{\kilo\meter\per\second}), and two of them had a co-temporal component with a plane-of-sky velocity of up to \SI[separate-uncertainty = true]{77(19)}{\kilo\meter\per\second}, starting near one of the loop footpoints. In one case, we measured a co-temporal intensity increase in the \ion{Si}{iv} line with IRIS that is associated with Doppler velocities down to \SI{-32}{\kilo\meter\per\second} and up to \SI{18}{\kilo\meter\per\second} along the line of sight. Finally, we measured cases of small-scale ($\approx$\SI{8e16}{\maxwell}) mixed-polarity field emergence and fluctuation near one of the loop footpoints.}
   {We conclude that the fast components on the plane-of-sky are consistent with a thermal transfer or supersonic plasma flows, while the slower component is consistent with plasma flows. A possible physical origin for these EUV emission enhancements would be magnetic reconnection driven by either a photospheric motion of the loop footpoints or by the reconnection of the loop with small-scale magnetic bipoles.}

    \titlerunning{Small-scale impulsive EUV emission enhancements along network loops}
    \authorrunning{Dolliou et al.}
   \maketitle
    \nolinenumbers

\section{Introduction}

The solar corona is heated to temperatures of multi-million kelvin. Magnetic reconnection \citep[][]{Pontin_2022} and waves \citep[][]{VanDoorsselaere2020} are thought to play a major role in the energy transfer and release from the photosphere to the corona. Imaging in the extreme-Ultraviolet (EUV) and X-rays shows that the heating is impulsive \citep[][]{Ugarte_2019,viall_survey_2017} and dominates at small scales \citep[][]{Crosby_1993,Hannah_2008,Berghmans_1998,Aschwanden_2000}. One of the main theories suggests that the shuffling and intertwisting of magnetic strands through photospheric motions leads to reconnection \citep[][]{Parker_1972} and energy release by a large number of small-scale events called nanoflares \citep[$\sim$\SI{e24}{\erg};][]{Parker1988}. While nanoflares were historically associated with magnetic reconnection with energy dissipation in current sheets, the term is now used to refer to any small-scale energy release event, regardless of its physical origin \citep[][]{Klimchuk_2015}. These origins include reconnection, a wave, or a mix of the two, such as wave-induced reconnection \citep[][]{Sukarmadji_2024}.

Coronal loop bundles in the quiet Sun (QS) are shorter ($\sim$ \SI{20}{\mega\meter}) and have a shorter lifetime (a few hours) than loops in active regions \citep[ARs; up to \SI{100}{\mega\meter} of length and multiple days of lifetime;][]{Reale2014}. We refer to these QS loops as "network loops" since they are rooted in the chromospheric network and their apparent lengths are comparable to that of a supergranule ($\sim$ \SI{30}{\mega\meter}). The lengths and lifetimes of these bundles of network loops are comparable to those of coronal bright points \citep[CBPs;][]{Madjarska_2019}. The latter are comparatively brighter in the EUV, with the more energetic ones showing signatures even in X-rays. Nevertheless, the denominations "CBPs" and "network loops" can refer to the same structures when the CBPs are anchored in the intergranular network. In the rest of the paper, we use the network loops and CBP terminology interchangeably. \cite{Doschek_2010} used data from the EUV Imaging Spectrometer \citep[EIS;][]{Culhane_2007} on board the Hinode spacecraft \citep[][]{Kosugi_2007} to estimate the peak temperature of one CBP to be \SI{1.5}{\mega\kelvin}. These loops are of particular interest as they are always present in the QS, with at least 250 bright points detectable at any given time on the whole solar disk \citep[][]{McIntosh_2005}. Thus, CBPs contribute significantly to the EUV emission in the QS. Furthermore, CBPs are often considered as scaled-down versions of ARs. Processes such as chromospheric evaporation and plasma cooling are more likely to be captured in full in CBPs by remote-sensing observations because their timescales in CBPs are shorter than in AR loops. For instance, the conduction cooling timescale \citep{Cargill_2004} of a large AR loop \citep[$n\sim$ \SI{1e10}{\per\centi\meter\tothe{3}}, $L\sim$ \SI{10e9}{\centi\meter}, $T\sim$ \SI{3e6}{\kelvin};][]{Reale2014} is equal to about two  hours, while that of a typical CBP ($n\sim$ \SI{5e9}{\per\centi\meter\tothe{3}}, $L\sim$ \SI{1e9}{\centi\meter}, $T\sim$ \SI{2e6}{\kelvin}) is equal to about \SI{100}{\second}.

It is challenging to determine observational evidence of the different heating processes that sustain the energy balance of CBPs. A combination of steady background (low-level) heating and a storm of more impulsive heating is likely required to sustain the CBPs \citep[][]{2013ApJ...768...32C}.  With regard to waves, EUV imaging can detect transverse oscillations \citep[][]{Lim_2023,Duckenfield_2018,Nakariakov_2021,Mandal_2022} and longitudinal slow magneto-acoustic waves \citep[][]{Prasad_2014,DeMortel_2009}. In magnetic reconnection, the current-sheet thickness can reach values down to the $\sim$ \SI{1}{\meter} scale for typical CBP parameters \citep[length $L\sim$\SI{e9}{\centi\meter}, Lundquist number $S \sim$ \SI{e13}{} -- \SI{e14}{};][]{Priest_1982}. Thus, the length scales of the current sheets  remain below the spatial resolutions of the currently best EUV imagers (down to $\sim$ \SI{200}{\kilo\meter}). Only indirect signatures can be measured, such as EUV emission of the surrounding cooling plasma \citep[][]{Chitta_2022} or moving intensity peaks in the EUV emission that have been interpreted as plasmoid jets \citep[][]{Duan_2025,nobrega_sivero_2025}. 

In recent years, high-resolution EUV imaging observations have revealed brightenings in CBPs and other small-scale loops \citep[][]{Tiwari2022,Mandal_2021,Winebarger_2013}. In particular, small (100 - \SI{4000}{\kilo\meter}) and short-lived (3 - \SI{200}{\second}) EUV brightenings \citep[][]{Berghmans2021,Narang_2025} are regularly detected with the Extreme Ultraviolet Imager \citep[EUI;][]{EUI_instrument} on board the Solar Orbiter mission \citep[][]{Muller,Zouganelis2020}. These brightening events, often called "campfires," have mainly been detected in the QS, but also in CBPs \citep[][]{Tiwari2022}. They have been interpreted as the signature of the plasma response to magnetic reconnection \citep[][]{Kahil_2022,Panesar_2021,Nelson_2024,chen2021} or to impulsively generated Alfv\'en waves \citep[][]{Kuniyoshi_2024}.

We report the detection of small-scale EUV emission enhancements along network loops. Periodic EUV emission enhancements along coronal loops have been investigated for decades in many structures and at different scales:  the pulsations caused by thermal instability and thermal non-equilibrium (TI-TNE) cycles in AR loops \citep[][]{Froment_2015,Froment_2017,Auchere_2018}, and quasi-periodic pulsations (QPPs) observed from large-scale flares \citep[][]{Nakariakov_2009} to small-scale EUV brightenings \citep[][]{Lim_2025}. Pulsations in EUV emission with similar properties as those reported in our work have also been found in CBPs \citep[][]{Matsumoto_2025}. The novelty of this work relies in the detection of a large number (about two dozen) of EUV emission enhancements per hour in three loop bundles and in the kinematic analysis of four of them with time-distance maps. This was possible through the high spatial ($\approx$ \SI{206}{\kilo\meter}) and temporal resolution (\SI{10}{\second} cadence) of the High Resolution EUV Imager (HRIEUV) at a time when Solar Orbiter was at 0.293 AU from the Sun. We also chose a multi-instrumental approach to investigate whether they show properties consistent with magnetic reconnection or with slow magneto-acoustic waves. Regarding the latter, we refer to features such as propagating disturbances \citep[PDs;][]{DeForest_Gurman_1998} with plane-of-sky (PoS) velocities close to the speed of sound when observed in different EUV channels \citep[][]{Kiddie_2012}. An important remark is that this observational property does not necessarily mean that the origins of PDs are slow magneto-acoustic waves. They also show faint and periodic blue-wing enhancements with spectroscopy \citep[][]{Tian_2011b}. Thus, their origin remains unclear, and they have been interpreted as signatures of either plasma flows or slow magneto-acoustic waves \citep[][]{Prasad_2014}.

This paper is organized as follows. In Sect. \ref{sec:observation} we introduce the data we used in this work, the instruments, and the main network loop bundles of interest. In Sect. \ref{sec:results} we show the main observational results, which we discuss in Sect. \ref{sec:discussion}. We summarize our result and draw our conclusions in Sect. \ref{sec:conclusions}.

\section{Observations}
\label{sec:observation}

\begin{figure*}
    \centering
    \includegraphics{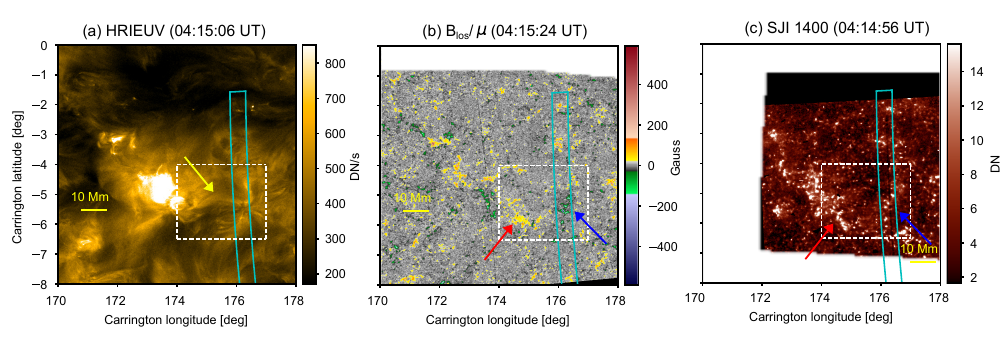}
    \caption{Observations of network loop bundles. HRIEUV (a), PHI-HRT (b), and IRIS SJI 1400 (c) images showing the region surrounding the network loops of interest. The images are those closest in time. The blue rectangle shows thed FOV of the IRIS rasters. The time displayed for SJI 1400 has been corrected to take the difference of \SI{354}{\second} in light time travel from the Sun to Earth and the Sun to Solar Orbiter into account. The white rectangle shows the FOV of Fig. \ref{fig:results:loop1_2_3_fov}a. The yellow arrow (a) highlights the loop bundle of interest seen in HRIEUV, and the red and blue arrows indicate the locations in the network in which it is assumed to be anchored.}
    \label{fig:observations:full_fov}
\end{figure*}

We used the 2023 April 10 observation of a QS region (from 03:30 to 09:30 UT) that was part of the Solar Orbiter Observing Plan \citep[SOOP;][]{Auchere_2020,Zouganelis2020} R\_BOTH\_HRES\_HCAD\_Nanoflares. The observation was coordinated between multiple instruments on board Solar Orbiter, including EUI and the Polarimetric and Heliosismic Imager \citep[PHI;][]{Solanki_2020}. The coordination also involved instruments from spaced-based observatories in Earth orbit, in particular, the Interface Region Imaging Spectrograph \citep[IRIS;][]{DePontieu2014}. On 2023 April 10, Solar Orbiter was at a distance of 0.293 AU from the Sun, which is one of the closest perihelia of the mission. This also resulted in a difference of $\approx$ \SI{354}{\second} between the light travel times from the Sun to Solar Orbiter and from the Sun to Earth. The heliographic longitude of Solar Orbiter with respect to Earth  was \SI{62.6}{\degree}, and the EUI FOV was centered \SI{-27.2}{\degree} away from the Sun center, in the Heliographic frame. We chose this specific observation sequence as it combines multi-instrumental observations over a long period (6 hours) with high spatial resolution and cadence for the instruments on board Solar Orbiter. In the next sections, we describe in detail the properties and the data-processing steps of the instruments (Sect. \ref{sec:obs:instruments}), and we present the three network loop bundles studied in this work (Sect. \ref{sec:obs:network_loops}). We also highlight one of the loop bundles as of particular interest because one of its footpoints is included in the field of view (FOV) of the IRIS spectrograph.

\subsection{Instruments}
\label{sec:obs:instruments}

\subsubsection{EUI on board Solar Orbiter}
\label{sec:obs:instruments:eui}

The instrument EUI is composed of the Full Sun Imager (FSI) and two high-resolution imagers (HRIs), named HRIEUV and HRILy$\alpha$. FSI includes two filters centered around \SI{304}{\angstrom} (\ion{He}{ii} \SI{303.78}{\angstrom}) and \SI{174}{\angstrom} (\ion{Fe}{X} \SI{174.53}{\angstrom}). We mainly used HRIEUV data. FSI was only used to co-register the HRIEUV dataset.

The HRIEUV filter has a passband centered around the \ion{Fe}{X} \SI{174.53}{\angstrom} line (forming at about \SI{1}{\mega\kelvin}) and also includes a contribution from lines emitted by plasma at TR temperatures. The part of the APS sensor that is sent to ground has 2048$\times$2048 pixels, with a pixel size equal to \SI{0.492}{\arcsec}, resulting in a \SI{1007.6}{\arcsec}$\times$\SI{1007.6}{\arcsec} FOV. At 0.293 AU from the Sun, the pixel size was equivalent to 
\SI{103}{\kilo\meter} on the solar atmosphere, given a re-projection on a sphere of radius 1.004 $R_\mathrm{sun}$. The spatial resolution, tested on flight, reaches the Nyquist limit of two pixels \citep[][]{Berghmans_2023}. The cadence was set to \SI{10}{\second} during the whole sequence.

We used level-2 (L2) FITS files from the EUI data release 6.0 \citep[][]{euidatarelease6}. The data-processing from level 1 (L1) to L2 included a dark field and a flat-field correction, the correction to the intensity for high-gain pixels, and the normalization of the intensity by the exposure time. The pointing information on the FSI metadata was also previously corrected with limb-fitting. We corrected the pointing information in the HRIEUV metadata by cross-correlating HRIEUV images with limb-fitted FSI images closest in time. The correlation was performed with the \software{Python} package \software{euispice\_coreg}\footnote{\url{https://github.com/adolliou/euispice_coreg}, consulted on 2025 September 1}, which is described in Appendix A of \cite{Dolliou_2024}. We also corrected the spacecraft jitter with a method inspired by the one described in Appendix A of \cite{Chitta_2022}: the HRIEUV dataset was first decomposed into one-minute sublists. Then, every image was cross-correlated with the first image of its sublist. As each sublist overlaps with the next by one image, the spacecraft jitter was corrected over the whole dataset. The images were then reprojected into a Carrington grid with longitude and latitude limits equal to [\SI{163.5}{\degree}, \SI{186}{\degree}] and [\SI{-11.5}{\degree}, \SI{11}{\degree}], respectively. The resulting 2680 $\times$ 2680 grid has a pixel size of \SI{0.0084}{\degree}, and the reprojection was performed on a sphere of radius $1.004 R_{\sun}$. Thus, the distance on the atmosphere associated with an HRIEUV pixel (\SI{103}{\kilo\meter}) was preserved.

\subsubsection{PHI on Solar Orbiter}
\label{sec:obs:instruments:phi}

The instrument PHI measures the full-vector magnetic field on the photosphere. It is composed of two separate telescopes: the High Resolution Telescope \citep[HRT;][]{Gandorfer_2018} and the Full Disk Telescope (FDT). We only used PHI-HRT to measure the photospheric magnetic field at the highest possible spatial and temporal resolution.

The instrument PHI-HRT measures the four Stokes parameters (I, U, Q, and V) at five spectral positions across the \ion{Fe}{I} \SI{617.3}{\nano\meter} line and at one position in the near continuum. The full photospheric magnetic field vector ($B_\mathrm{mag}$, $B_\mathrm{inc}$, and $B_\mathrm{azm}$) is computed with a radiative transfer equations (RTE) inversion, which assumes that light propagates under Milne-Eddington atmospheric conditions \citep{2007A&A...462.1137O}. A \SI{180}{\degree} ambiguity still exists on the $B_\mathrm{azm}$ component of the field \citep[see][for more details]{Metcalf_2006}. In this work we used images of the line-of-sight component $B_\mathrm{los}$. We divided $B_\mathrm{los}$ by the cosine of the Heliocentric angle $\mu$ at each pixel to obtain an estimation of the radial component of the field $B_\mathrm{rad}$, assuming the magnetic field is predominantly radial. 

In the observation sequence we used, the images from PHI-HRT are cropped to 1024$\times$1024 square pixels of size \SI{0.5}{\arcsec}, resulting in a \SI{512}{\arcsec}$\times$\SI{512}{\arcsec} FOV. On 2023 April 10, the pixel size was equivalent to \SI{105}{\km} per pixel at disk center on the photosphere. The spatial resolution is designed to be at the 2 pixels Nyquist limit. The cadence was set to \SI{1}{\min} during the six hours sequence.

We used L2 FITS files from the third PHI-HRT data release\footnote{\url{https://www.mps.mpg.de/solar-physics/solar-orbiter-phi/data-releases}, consulted on 2025 September 1.}. The full data-processing pipeline \citep[based on][]{sinjan_2022,Kahil_2023,2024-bailen} is available with the Python package \software{hrt\_pipeline}\footnote{\url{https://github.com/dcalc/hrt_pipeline}}. In addition, we applied the post-processing step described in \cite{Calchetti_2023}: a pixel-by-pixel linear interpolation of the spectro-polarimetric data between the consecutive frames, to ensure that all data are given at the same time. The pointing information in the PHI-HRT metadata has been corrected using the line-of-sight magnetic field ($B_\mathrm{los}$) maps with the ones closest in times from the Heliosismic and Magnetic Imager \citep[HMI;][]{Scherrer_2012,2012SoPh..275..229S}, on board the Solar Dynamics Observatory \citep[SDO;][]{Pesnell2012}. The PHI-HRT images have been remapped into the HMI coordinates, and the co-alignment in performed with a cross-correlation method. Finally, the PHI-HRT images are re-projected to Carrington frame with longitude and latitude limits equal to [\SI{167}{\degree}, \SI{180.1}{\degree}] and [\SI{-11.0}{\degree}, \SI{2.1}{\degree}] respectively. The resulting 1496$\times$1496 Carrington grid has a pixel size of \SI{0.0084}{\degree} to preserve PHI-HRT pixel size on the photosphere. The re-projection is performed on a sphere of radius $1.000 R_{\sun}$.

\subsubsection{IRIS}
\label{sec:obs:instruments:iris}

The instrument IRIS is an imaging spectrograph that observes in the Far UV (FUV), specifically in the  \SI{1331}{}-\SI{1358}{\angstrom} and \SI{1390}{}-\SI{1407}{\angstrom} ranges; in the Near UV (NUV), specifically in the \SI{2782}{\angstrom}-\SI{2834}{\angstrom} range; and with the slit jaw imager (SJI) covering four bandpasses in the FUV and in the NUV. On 2023 April 10, we used the SJI bandpass centered at \SI{1400}{\angstrom}. The main FUV lines of interest measured by the spectrograph are the followings : \ion{C}{ii} \SI{1335.66}{\angstrom} + \SI{1335.70}{\angstrom} ($\log{T}\mathrm{[K]} = 4.5$) ; \ion{Si}{iv} \SI{1393.76}{\angstrom} ($\log{T}\mathrm{[K]} = 4.8$) ; \ion{Si}{iv} \SI{1402.77}{\angstrom} ($\log{T}\mathrm{[K]} = 4.8$) and the NUV line \ion{Mg}{ii} k \SI{2796.35}{\angstrom}. For the FUV lines, the log T value shown in brackets corresponds to the line formation temperature under equilibrium conditions.

The spectrograph was set to a raster mode with a slit of length \SI{120}{\arcsec} and width \SI{0.33}{\arcsec}. The raster had eight exposures, with a step size of \SI{1}{\arcsec} and a FOV of \SI{7}{\arcsec}$\times$\SI{120}{\arcsec}. The step size corresponds to $\approx$\SI{720}{\kilo\meter} on the Sun's atmosphere, at 1.0008 $R_\sun$ from the Sun's center. The exposure time was equal to \SI{8}{\second}, and the full raster cadence to \SI{70}{\second}. The SJI images had a FOV of \SI{120}{}$\times$\SI{120}{\arcsec}, with a pixel size of \SI{0.17}{\arcsec}. The latter is equivalent to a length of \SI{123}{\kilo\meter} on the Sun's atmosphere, at 1.0008 $R_\sun$ from the Sun's center.

The L2 data are obtained from the heliophysics events knowledgebase coverage registery (HCR)\footnote{\url{https://www.lmsal.com/hek/hcr}, consulted on 2025 September 2.}. The L1 to L2 data-processing steps included a dark and flat field correction ; a geometric and wavelength calibration. We counted for the residual wavelength shift in the pipeline correction. To do so, we averaged the spectra over a quiet region in the upper part of the raster FOV. We then fitted the \ion{O}{i} \SI{1355.60}{\angstrom} line with a single Gaussian, and computed the Doppler velocity at each time step. Finally, we shifted the wavelength array by the same constant value of \SI{0.0356}{\angstrom} for all rasters, so that the median of the \ion{O}{i} Doppler velocities over the whole sequence get shifted to \SI{0}{\kilo\meter\per\second}.

We also co-registered the SJI 1400 and 2796 images with images from the Atmospheric Imaging Assembly \citep[AIA;][]{Lemen2012}, on board SDO. The filter used was the one centered around \SI{1700}{\angstrom}, which mainly images the chromosphere. The AIA dataset had been processed from L1 to L1.5 with the \software{aiapy} python package \citep{Barnes2020}. The IRIS spectrograph has been also visually co-aligned to the SJI 2796 dataset. To do so, the position of the slit in the metadata of the spectrograph FITS files is set to match the slit visible in SJI images. In addition, the position of fiducial marks on the \ion{Mg}{ii} h/k spectrum are set to match the ones visible on SJI images. In this work the analysis of IRIS data was performed with the help of the \software{irispy} python package\footnote{\url{https://github.com/LM-SAL/irispy}, consulted on 2025 September 2.}.

\subsection{Network loops of interest}
\label{sec:obs:network_loops}

The main network loop of interest, called "loop bundle 1" (yellow arrow in Fig. \ref{fig:observations:full_fov}a), and the surrounding QS region are displayed in an HRIEUV, PHI-HRT and a SJI 1400 image. The FOV of the IRIS spectrograph raster map is also displayed. The opposite-polarity footpoints, visible in PHI-HRT, are rooted in the chromospheric network, visible in SJI 1400 (blue and red arrows in Figs. \ref{fig:observations:full_fov}b and \ref{fig:observations:full_fov}c). The FOV of the IRIS spectrograph raster includes the right footpoint of the loop with a negative polarity. The HRIEUV images show the EUV emission from the plasma located on the TR and the coronal part of the loop. The loop bundle itself has an apparent length of about \SI{20}{\mega\meter}, which is comparable to the supergranulation scale \citep[$\approx$ \SI{30}{\mega\meter};][]{Rieutord_2010}, and a lifetime of about one hour (from 04:00 UT to 05:15 UT). After 05:15 UT, the loop brightens significantly and its morphology, as seen in HRIEUV, drastically changes. The cause could be a change in its thermal profile or a reconfiguration of the field lines following a major reconnection event. In this work, we focus on the time interval when the large-scale shape of the loop bundle appears stable in HRIEUV, from 04:00 UT to 05:15 UT. A movie showing loop bundle 1 over this time interval is available online ("loop\_bundle1.mp4"). We also included in this work two other loop bundles of similar apparent lengths as loop bundle 1, called loop bundles 2 and 3 (see Fig. \ref{fig:results:loop1_2_3_fov}). The time ranges of loop bundles 2 and 3 are from 06:55 to 08:30 UT and from 03:35 to 06:00 UT, respectively.

\section{Results}
\label{sec:results}

In the following, we turn to the analysis of the observations that is mainly based on the investigation of small-scale enhancements of EUV emission in network loops. Section \ref{sec:results:impulsive_qpp} introduces the detection of these impulsive EUV emission enhancements. We estimate the PoS velocities for four of them in Sect. \ref{sec:results:apparent_v}. In Sect. \ref{sec:results:doppler_chromosphere} we verify with IRIS that one of the EUV emission enhancements is associated with bi-directional plasma flows in the TR. In Sect. \ref{sec:results:magnetic} we look for the photospheric drivers of the EUV emission enhancements.

\subsection{Detection of small-scale impulsive EUV emission enhancements along network loop bundles}
\label{sec:results:impulsive_qpp}

\
\begin{figure*}
\centering
\includegraphics{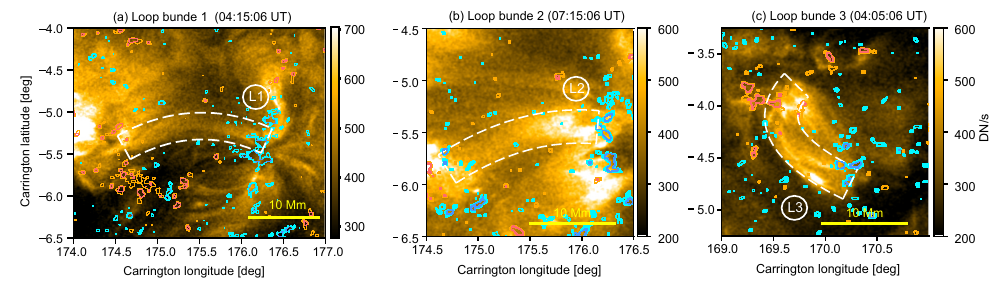}
\caption{Location of the slits for the three loop bundles. HRIEUV images with a FOV centered around (a) loop bundle 1 at 04:15:06 UT, (b) loop bundle 2 at 07:15:06 UT, (c) and loop bundle 3 at 04:05:06 UT. The $B_\mathrm{los}/\mu$ values measured by PHI-HRT closest in time are shown as dark orange ($B_\mathrm{los}/\mu>$ \SI{50}{\gauss}), light orange ($\SI{30}{\gauss} > B_\mathrm{los}/\mu \geq \SI{50}{\gauss}$), light blue ($\SI{-30}{\gauss} > B_\mathrm{los}/\mu > \SI{-50}{\gauss}$), and dark blue contours ($\SI{-50}{\gauss} > B_\mathrm{los}/\mu$). The slits along which the time-distance maps are computed are shown as dashed white lines. The slits are named L1, L2, and L3 for loop bundles 1, 2, and 3, respectively. The yellow bar in the bottom right corner of each panel represents \SI{10}{\mega\meter} and is shown as a reference. The associated movie for L1 is available online.}
\label{fig:results:loop1_2_3_fov}
\end{figure*}

\begin{figure}
    \centering
    \includegraphics[scale=0.95]{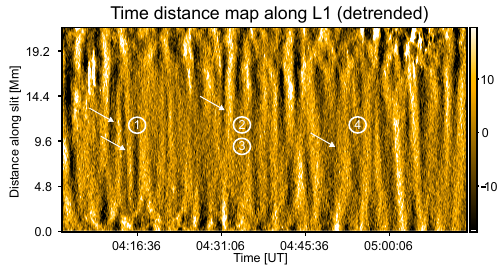}
    \caption{Time-distance map along slit L1 covering network loop bundle 1 (Fig. \ref{fig:results:loop1_2_3_fov}a). In this figure only, the signal is detrended for visualization purposes. The white arrows indicate the four EUV emission enhancements investigated in Sect. \ref{sec:results:apparent_v}. The numbers 1 to 4 are the labels of the slits over which the apparent velocity of these EUV emission enhancements are measured (Fig. \ref{fig:results:all_slits_fov_oneimage}). }
    \label{fig:results:loop1_stackplot_detrended}
\end{figure}

As a first step we measure the EUV emission enhancements along the three loop bundles. To do so, we place slits along the loops, which are displayed on an HRIEUV image in Fig. \ref{fig:results:loop1_2_3_fov}. The slits, called L1 to L3, are designed to cover the full width of the loop bundles 1 to 3. The shapes of the slit axis are piecewise cubic polynomials. We computed, at each point along the slit, the vector normal to the slit axis with a norm equal to the slit width. In the following, the time-distance maps are computed by averaging the intensities over these normal vectors at each point of the slit axis.

Multiple impulsive EUV emission enhancements along L1 (Fig. \ref{fig:results:loop1_stackplot_detrended}) are detected from 04:00 to 05:15 UT, with four of them being highlighted by white arrows. Some of the EUV emissions enhancements extend from one footpoint to the other. Two examples of this case are the enhancements  highlighted by white arrows before 04:16:36 UT. The data used in Fig. \ref{fig:results:loop1_stackplot_detrended} has been detrended by subtracting a \SI{210}{\second} running average to each spatial location. We note that the detrending was performed for visualization purposes only. No detrending is applied to the data in the following analysis, including in Sect. \ref{sec:results:apparent_v}. The time-distance map along L1 with the original data is given in Fig. \ref{fig:results:loop1_2_3_stackplots}, along with the time-distance maps along L2 and L3. We observe impulsive EUV emission enhancements in all of the three loop bundles during their respective time range. These emission enhancements are separated by timescales in the order of a few minutes. We counted for instance 6 to 8 events in the time intervals from 04:31 to 04:45 UT in Fig. \ref{fig:results:loop1_2_3_stackplots}a, and from 07:25 to 07:40 UT in Fig. \ref{fig:results:loop1_2_3_stackplots}b. We refer to Sect. \ref{sec:discussion:photospheric_driving} for a more thorough discussion on the possible photospheric drivers.  

In the next section, we choose a case-by-case approach to create a slit adapted for each EUV emission enhancement. The large width of the slit in Fig. \ref{fig:results:loop1_2_3_fov}a can induce errors when the PoS velocities of the EUV emission enhancements are measured. These errors on the PoS velocity can be caused by multiple EUV emission enhancements at the same time or by motions across the slit.

\subsection{Plane-of-sky velocities of the intensity peaks along the loop bundles}
\label{sec:results:apparent_v}

\begin{figure*}
    \centering
    \includegraphics{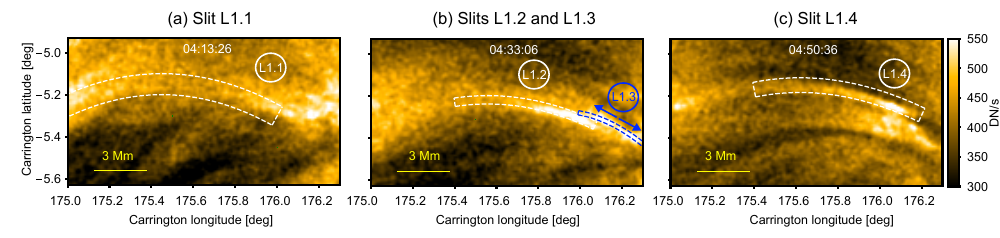}
    \caption{HRIEUV images zoomed-in to the right footpoint of loop bundle 1, showing (a) slit L1.1 at 04:13:26 UT, (b) slits L1.2 and L1.3 at 04:33:06 UT, and (c) slit L1.4 at 04:50:36 UT. The slits are displayed as either white or blue dotted lines. The blue arrows above slit 3 highlight am apparent bi-directional motion of intensity peaks. The associated movies are available online.}
    \label{fig:results:all_slits_fov_oneimage}
\end{figure*}

\begin{figure*}
    \centering
    \includegraphics{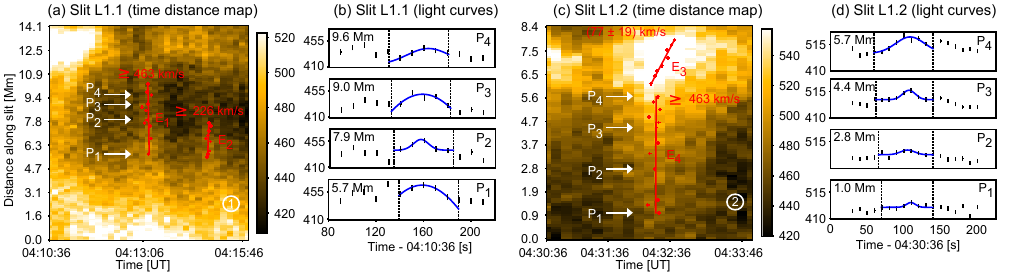}
    \caption{Time-distance maps (a and c) along slits L1.1 and L1.2 (Figs. \ref{fig:results:all_slits_fov_oneimage}a and \ref{fig:results:all_slits_fov_oneimage}b). The number of each slit is shown as a white circle in the bottom-right side. The intensity motions of interest are the events $E_1$ to $E_4$. The red crosses show the times of the intensity peaks associated with the events at different locations. The red line represents the computed PoS velocity on the time-distance map for each event. Panels (b) and (d) show four examples of light curves at distinct locations (noted $P_1$ to $P_4$) at the times of $E_1$ and $E_4$, respectively. The vertical dotted lines represent the time intervals that include the intensity peak. A Gaussian function (in blue) is fitted to estimate the central time of the intensity peak (see Appendix \ref{sec:annex:measurement_apparent_v} for more details). The events $E_1$, $E_2$ and $E_4$ seem to appear almost instantaneously along the slit, within the limitations due to the HRIEUV temporal resolution. In that case, we only show the highest resolvable PoS velocity as a lower limit. The intensity values are in DN/s. }
        \label{fig:results:v_apparente_slit_124_world}
\end{figure*}

\begin{figure}
    \centering
    \includegraphics{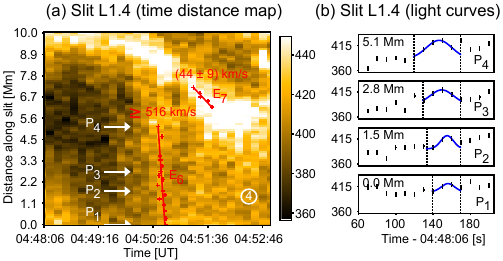}
    \caption{Same as Fig. \ref{fig:results:v_apparente_slit_124_world}, but for events $E_6$ and $E_7$ measured with slit L1.4 (Fig. \ref{fig:results:all_slits_fov_oneimage}c). Event $E_6$ seems to appear almost instantaneously, within the limitations due to the HRIEUV temporal resolution.}
    \label{fig:results:v_apparente_slit_5_world}
\end{figure}

The aim of this section is to measure the PoS velocities of the HRIEUV intensity peaks associated with four EUV emission enhancements along loop bundle 1, highlighted by the white arrows in Fig \ref{fig:results:loop1_stackplot_detrended}.

We first created the slits to measure the PoS velocities of the EUV emission enhancements along the loop. These slits are denominated from L1.1 to L1.4, and their numbering follows the time order of when the EUV emission enhancements are measured. They are narrower than L1 in order to include more precisely the strands visible in HRIEUV. They are also less extended than L1, as they only include the region where EUV emission enhancements are clearly detected. The slits are displayed in images of HRIEUV at the times of their respective EUV emission enhancements (Fig. \ref{fig:results:all_slits_fov_oneimage}). The full movies showing slits L1.1 to 1.4 are available online (slit\_L1\_1.mp4, slits\_L1\_2\_and\_L1\_3.mp4, and slit\_L1\_4.mp4). In the following, we describe the main take-away points from Fig. \ref{fig:results:all_slits_fov_oneimage} and the movies. Slits L1.1 and L1.2 are placed along strands with an intensity peak visible in HRIEUV on their right footpoint (\SI{176.0}{\degree} of longitude) from 04:10 to 04:12 UT and from 04:32:46 to 04:33:46 UT, respectively. On slit L1.4 (Fig. \ref{fig:results:all_slits_fov_oneimage}c), the intensity first increases along the whole strand, followed by an intensity peak motion from the right toward the left footpoint, starting at 04:50:23 UT on a longitude of \SI{176.1}{\degree}. This brightening disappeared when it reached a longitude of \SI{175.8}{\degree}, at 04:52:26 UT. Unlike the others, slit L1.3 does not capture an EUV emission enhancement along the loop. Instead, it captures bi-directional intensity motions near the right footpoint of loop bundle 1 (indicated by two blue arrows in Fig. \ref{fig:results:all_slits_fov_oneimage}b). The latter is co-temporal with the EUV emission enhancement measured with slit 2.

We measured the PoS velocities of the EUV emission enhancements along the slits. The intensities were averaged at each time step over the widths of slits L1.1 to L1.4 to obtain the time-distance maps (Figs. \ref{fig:results:v_apparente_slit_124_world}, \ref{fig:results:v_apparente_slit_5_world}, and \ref{fig:annex:v_apparente_slit_3_world}).  By convention, the distance increases while getting closer to the right footpoint of loop bundle 1. In the following, we define as events the intensity peaks in time-distance maps that either propagate along the slit with a given PoS velocity or that appear instantaneously, with respect to the limitations due to the HRIEUV cadence of \SI{10}{\second}. We distinguished the names "events" from the "EUV emission enhancement," as the EUV emission enhancements measured with slits L1.2 and L1.4 are associated with more than one event co-temporally. Each "event" detected on the time maps of slits L1.1 to L1.4 was given a name from $E_1$ to $E_7$. To compute their PoS velocity, we identified the  intensity peaks of each event on the light curves at equidistant locations along the slits. Examples of these light curves are given in Figs. \ref{fig:results:v_apparente_slit_124_world}b, \ref{fig:results:v_apparente_slit_124_world}d, and \ref{fig:results:v_apparente_slit_5_world}b for events $E_1$, $E_4$, and $E_6$, respectively. The intensity peaks were fitted with a Gaussian functions. The values and the uncertainties of the central times of the Gaussian functions were used to compute the PoS velocities of the events (see Appendix \ref{sec:annex:measurement_apparent_v} for more details).

The results of the PoS velocity computation for each event are shown in Figs. \ref{fig:results:v_apparente_slit_124_world} and \ref{fig:results:v_apparente_slit_5_world} and in Table \ref{table:events}. The highest PoS velocity that could be resolved was the one of $E_3$ at \SI[separate-uncertainty = true]{77(19)}{\kilo\meter\per\second}. With regards to the events $E_1$, $E_2$, $E_4$ and $E_6$, they seem to appear almost instantaneously along the slit, under the limitations due to the temporal cadence of HRIEUV (\SI{10}{\second}). Figures \ref{fig:results:v_apparente_slit_124_world}b and \ref{fig:results:v_apparente_slit_124_world}\ref{fig:results:v_apparente_slit_5_world} show two EUV emission enhancements with two distinct co-temporal events : an event appearing almost instantaneously ($E_4$ and $E_6$) ; and an event with a resolved PoS velocity equal to \SI{77}{\kilo\meter\per\second} ($E_3$) or \SI{44}{\kilo\meter\per\second} ($E_7$). The time-distance map corresponding to slit L1.3 is displayed in Fig. \ref{fig:annex:v_apparente_slit_3_world}. It confirms the existence of bi-directional intensity motions of about \SI{21}{\kilo\meter\per\second} near the footpoint of loop bundle 1, at the time of the EUV emission enhancement measured with slit L1.2.

For slits L1.1 and L1.2, the strand could barely be identified, so we verified that $E_1$, $E_2$, $E_3$ and $E_4$ (Figs. \ref{fig:results:v_apparente_slit_124_world}a and \ref{fig:results:v_apparente_slit_124_world}b) did not originate from background emission below or above loop bundle 1 along the line of sight. To do so, we spatially shifted slits L1.1 and L1.2 to move them away from the loop bundle (Figs. \ref{fig:annex:background_slit_34}a and \ref{fig:annex:background_slit_34}c), and computed the time-distance maps. The events could not be identified in these "background" time-distance maps (Figs. \ref{fig:annex:background_slit_34}b and \ref{fig:annex:background_slit_34}d). Thus, they do not originate from the background emission.

Finally, we also estimated the intensity peaks compared to the background for the events $E_4$, $E_6$ and $E_7$ (see Appendix \ref{sec:annex:measurement_apparent_v} for the method). Our results showed that the three events reach up to 14\%, 12\%, and 18\% intensity peak above the background level, respectively. This highlights the importance of the high spatial resolution in order to detect a large number of the EUV emission enhancements.  

\subsection{Chromospheric and lower TR Doppler velocity}
\label{sec:results:doppler_chromosphere}
\begin{figure*}
    \centering
    \includegraphics[scale=0.96]{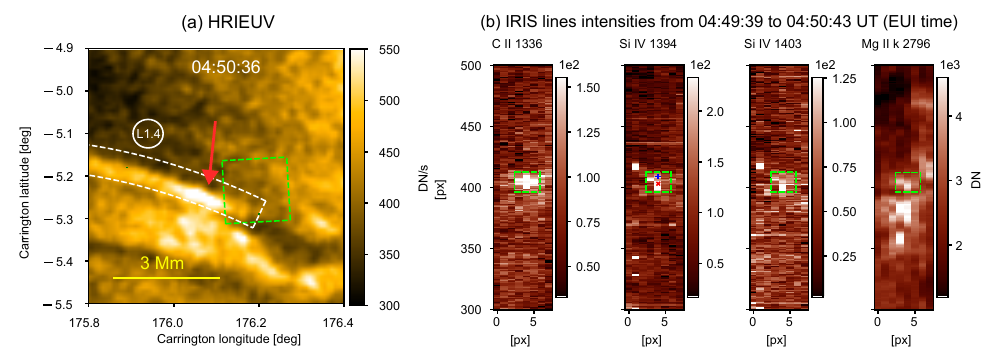}
    \caption{Chromospheric and TR line intensities at the time of event $E_7$ (Fig. \ref{fig:results:v_apparente_slit_5_world}d). Left: Image of HRIEUV (04:50:36 UT) showing slit L1.4 (white dotted region) and the intensity peak associated with $E_7$ (indicated by a red arrow). Right : Spectrally summed images of four UV lines over [-110, 110] \SI{}{\kilo\meter\per\second} measured with IRIS. The HRIEUV and IRIS images are zoomed-in around the locations of $E_7$ and slit L1.4. The rectangles in green correspond to the same spatial region in Carrington coordinates (a) and in IRIS pixels coordinates (b). The spectra in Fig. \ref{fig:results:iris_spec_event8} are spatially averaged over the region  enclosed by the rectangle in green shown in the HRIEUV and IRIS images. The blue cross and the red star displayed in the \ion{Si}{iv} \SI{1394}{\angstrom} images are the pixels over which the spectra at 04:50:11 UT are displayed in Fig. \ref{fig:annex:iris_spec_event8_v2_specific_points}. }
    \label{fig:results:iris_FOV_spec}
\end{figure*}

\begin{figure*}
    \centering
    \includegraphics[scale=0.97]{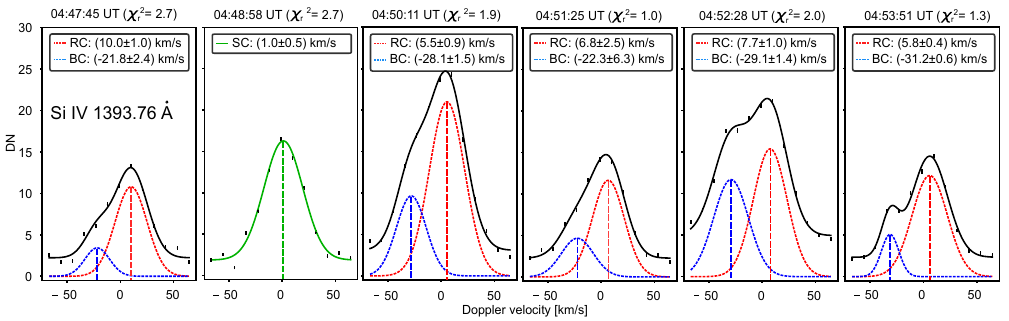}
    \caption{Spectra averaged over the region enclosed by a rectangle in green displayed in Fig. \ref{fig:results:iris_FOV_spec}, centered around \ion{Si}{iv} \SI{1393.76}{\angstrom}. The spectra are displayed for six times around the apparent starting time of $E_7$ at 04:50 UT. We apply a fitting with a single Gaussian to the spectrum at 04:48:58 UT called single component (SC) ; and a fitting with two Gaussian functions to the other spectra, called the red (RC) and the blue component (BC). The resulting velocity and uncertainty is written for each component in the legends. The vertical dashed lines indicate the locations in Doppler velocities of the peaks for the two components. The indicated times are the central times of the IRIS fast rasters. The $\chi_\mathrm{r}^2$ value resulting from each fit is indicated next to the time in UT for each spectrum.}  \label{fig:results:iris_spec_event8}
\end{figure*}

\begin{figure}
    \centering
    \includegraphics{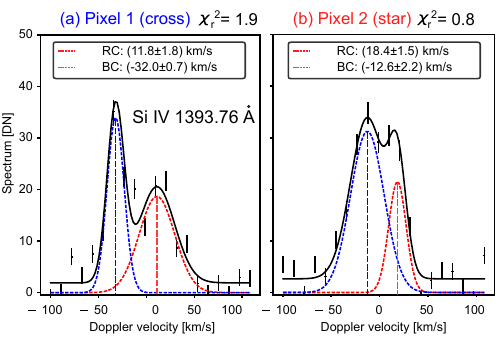}
    \caption{Spectra of \ion{Si}{iv} \SI{1393.76}{\angstrom} at 04:50:11 UT in pixel 1 (a) and pixel 2 (b), indicated by a blue cross and a red star in the \ion{Si}{iv} \SI{1393.76}{\angstrom} intensity map (Fig. \ref{fig:results:iris_FOV_spec}). The vertical dashed lines indicate the locations in Doppler velocities of the peaks for the two components.}
    \label{fig:annex:iris_spec_event8_v2_specific_points}
\end{figure}

In this section we aim to verify whether the EUV emission enhancements seen in HRIEUV are associated with any plasma flows in the lower part of the loop bundle. To do so, we present the measurements of intensity peaks and Doppler shifts with IRIS on UV lines emitted by plasma from chromospheric to lower TR temperatures. The IRIS spectra we consider in this section are taken in the region of the right footpoint of loop bundle 1, and at the time of the EUV emission enhancement measured with slit L1.4. This emission enhancement was the only among the four studied in Sect. \ref{sec:results:apparent_v} that could be associated with a clear intensity increase in the IRIS lines.

We first identify a TR emission increase at the loop footpoint, at the time and within \SI{1}{\mega\meter} of the EUV emission enhancement measured with slit 4 (Fig. \ref{fig:results:iris_FOV_spec}a). The small elongated bright region seen in HRIEUV along slit L1.4 corresponds to $E_7$. At the same time, we measured with IRIS an intensity peak in four UV lines (Fig. \ref{fig:results:iris_FOV_spec}b). The intensities of the lower TR lines (\ion{C}{ii} and the two \ion{Si}{iv} lines) peak at the same time as $E_7$ (04:50 UT), near the right footpoint of loop bundle 1.

We measured the spectra of the \ion{Si}{iv} \SI{1393.76}{\angstrom} line at six times from 0:47:45 UT to 04:53:31 UT (Fig. \ref{fig:results:iris_spec_event8}). The spectra were spatially averaged over the region enclosed by a rectangle in green in Fig. \ref{fig:results:iris_FOV_spec}. We used Gaussian functions to model the intensity peaks of the line. We tested whether the spectra in Fig. \ref{fig:results:iris_spec_event8} are best modeled by a single or a double Gaussian functions. To do so, we computed the values of the reduced $\chi^2$ for both models, defined as $\chi_\mathrm{r}^2 = \left(\sum_i \left(r_i/\sigma_i \right)^2\right)/(N - K)$, with $r_i$ and $\sigma_i$ being  the residue of the fit and the uncertainty, respectively, on the data at each point $i$; $N$ the total number of points ; and $K$ the number of fitted parameters (four for the single Gaussian, seven for the double Gaussian). We then selected the model with the $\chi_\mathrm{r}^2$ value closest to unity. The results showed that the double Gaussian was the better model for most spectra (e.g., $\chi_\mathrm{r}^2 = 1.9$ compared to $7.2$ at 04:50:11 UT), with the exception of the one at 04:48:58 UT which is better modeled by a single Gaussian ($\chi_\mathrm{r}^2 = 2.7$ compared to $3.8$).

For all spectra where we fit two Gaussian functions, we called the two functions the blue component (BC) and the red component (RC). For the fitting, we set the minimal and maximal bounds of the peak location to [-45, 0] and [0, 45] (in \SI{}{\kilo\meter\per\second}) for BC and RC, respectively; and of the maximal amplitude to [0, 50] and [5, 50] (in \SI{}{\dn}). For both components, the minimal bound for the half width $w$ at $1/e$ of the maximal amplitude is set to $w_\mathrm{th}=$ \SI{6.1}{\kilo\meter\per\second}, which is the thermal half width for the \ion{Si}{iv} line. The maximal bound for $w$ is set to \SI{27}{\kilo\meter\per\second}.

The results shows that between 04:48:58 UT and 04:50:11 UT, the spectrum switch from being best modeled by a single to a double Gaussian functions, and has intensity peak values increasing from \SI{15}{\dn} to \SI{25}{\dn}. This time corresponds to when E$_7$ is first detected in HRIEUV. The Doppler velocities reach up to \SI{-29}{\kilo\meter\per\second} and \SI{7.7}{\kilo\meter\per\second} at 04:52:28 UT for the BC and RC, respectively. Finally, we also examined whether faster flows could be detected on the individual pixels within the spatial region over which the spectrum in Fig. \ref{fig:results:iris_spec_event8} was averaged. Among these pixels, we highlight two extreme cases (Fig. \ref{fig:annex:iris_spec_event8_v2_specific_points}): an upflow reaching up to \SI{32}{\kilo\meter\per\second} (pixel 1) and a downflow reaching up to \SI{18}{\kilo\meter\per\second} (pixel 2). The other pixels show a wide range of Doppler velocities ranging between these two values. 

\subsection{Small-scale mixed-polarity magnetic structures in the right footpoint}
\label{sec:results:magnetic}

\begin{figure*}
\centering
    \includegraphics{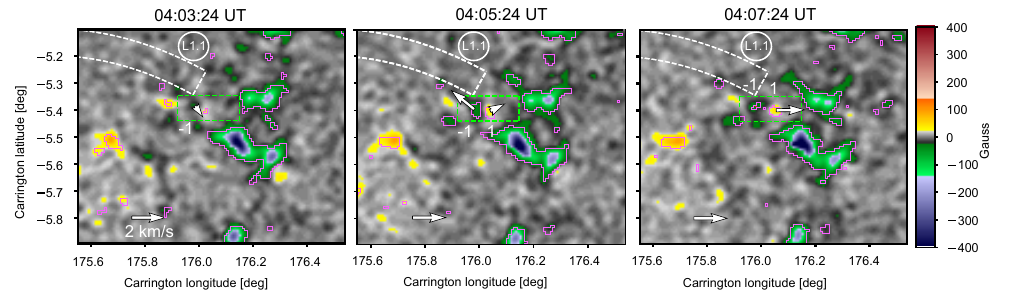}
\caption{Images of $B_\mathrm{los}/\mu$ obtained from PHI-HRT at times close to the EUV emission enhancement measured with slit L1.1 (Fig. \ref{fig:results:all_slits_fov_oneimage}a). The FOV is zoomed in to the right footpoint of the loop bundle 1. The arrows show the photospheric velocity vectors for magnetic structures of interest detected with SOFT (see Sect. \ref{sec:results:magnetic} for more details). The length of the arrows depends on the velocity norm, and an arrow representing \SI{2}{\kilo\meter\per\second} is shown as reference in the lower left part of the plot. We show two structures of interest named 1 and -1. The dotted green rectangle shows the region in which the signed flux is measured (Fig. \ref{fig:results:Bradflux_34_56}a). The pink contours indicate the  $4\sigma$ confidence level computed from the noise on the Stokes-V parameter (Appendix \ref{sec:annex:Blos_confint}).}
\label{fig:results:slit1_bmovement}
\end{figure*}

\begin{figure*}
\centering
    \includegraphics{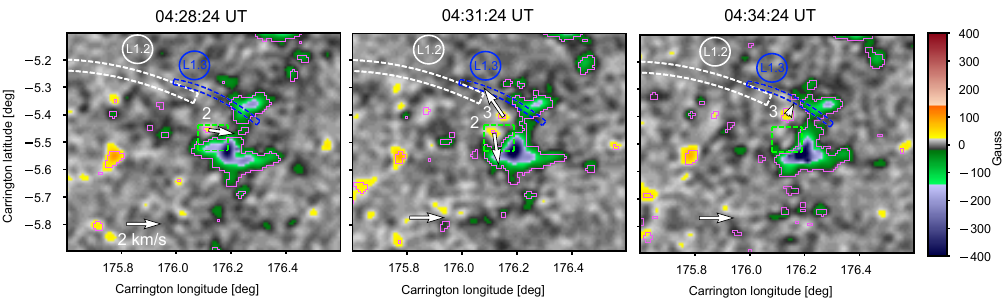}
    \caption{Same as Fig. \ref{fig:results:slit1_bmovement}, but for the EUV emission enhancement measured with slit L1.2 (dotted white region) and the bidirectional intensity motion measured with slit L1.3 (dotted blue region; see also Fig. \ref{fig:results:all_slits_fov_oneimage}b.}
    \label{fig:results:slit2_bmovement}
\end{figure*}

\begin{figure}
    \centering
    \includegraphics{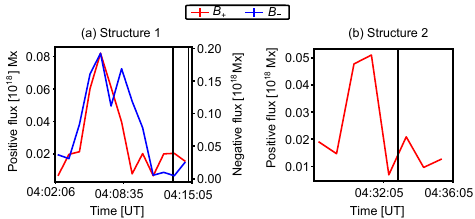}
    \caption{Positive ($B_{\mathrm{+}}$) and negative ($B_{\mathrm{-}}$) magnetic flux spatially summed over the region enclosed by the green rectangle displayed in Figs. \ref{fig:results:slit1_bmovement} and \ref{fig:results:slit2_bmovement}. The selected pixels of each region are those above the $3\sigma$ Stokes-V noise threshold (Appendix \ref{sec:annex:Blos_confint}). The vertical black lines indicate the time when EUV emission enhancements are detected with HRIEUV: $E_1$ and $E_2$ for (a) ; $E_4$ for (b).}
    \label{fig:results:Bradflux_34_56}
\end{figure}

The aim of this section is to look for signatures of magnetic drivers for the EUV emission enhancements detected in HRIEUV at the right footpoint of loop bundle 1. In particular, we focus on the signatures of small-scale flux emergence and cancellations. Figures \ref{fig:results:slit1_bmovement} and \ref{fig:results:slit2_bmovement} show $B_\mathrm{los}/\mu$ images measured by PHI-HRT at times close to the EUV emission enhancement measured in slit L1.1 (04:13 UT; Fig. \ref{fig:results:v_apparente_slit_124_world}a) and in slit L1.2 (04:33 UT; Fig. \ref{fig:results:v_apparente_slit_124_world}b), respectively. Here, $\mu$ refers to the cosine of the Heliocentric angle. The pink contours in the figure mark the pixels with maximal Stokes-V value four times above the noise level $\sigma$, the noise being defined as the standard deviation of the Stokes-V values over all pixels of the FOV at the near continuum wavelength, where noise dominates. Further details on the method used to compute the confidence levels are given in Appendix \ref{sec:annex:Blos_confint}. All $B_\mathrm{los}/\mu$ images show the existence of multiple small-scale magnetic structures with mixed polarities above the $4\sigma$ Stokes-V noise threshold near the negative-dominated polarity right footpoint of loop bundle 1 (Figs. \ref{fig:results:slit1_bmovement} and \ref{fig:results:slit2_bmovement}). These small-scale structures are present during the whole time interval over which loop bundle 1 is studied. For instance, in Appendix \ref{sec:annex:magnetic_slit5} we report their detection in large number above noise level at the times of the EUV emission enhancement measured with slit L1.4. They are similar to the mixed-polarity fields previously detected in the footpoints of AR loops \citep[][]{Chitta_2017} with the balloon-born SUNRISE observatory \citep[][]{Solanki_2010,Solanki_2017} ; and to the small-scale magnetic structures in the QS previously detected with PHI-HRT \citep[][]{Chitta_2023}.

In the following, we focus on four structures of interest that appeared minutes before the EUV emission enhancements measured with slits L1.1 and L1.2 (Figs. \ref{fig:results:slit1_bmovement} and \ref{fig:results:slit2_bmovement}). These structures are of interest because of the following properties: they are above the $4\sigma$ Stokes-V noise threshold, and their flux emerges and fluctuates near the footpoint of loop bundle 1 or near the starting location of events seen in HRIEUV. These structures are denominated from 1 to 3. Those with a negative polarity are named with a minus sign (e.g., -1 in Fig. \ref{fig:results:slit1_bmovement}). Their photospheric velocity vectors were estimated with the Solar Feature Tracking algorithm \citep[SoFT;][]{Berretti_2025}, given a detection threshold of \SI{21}{\gauss}.  We also measured the evolution of the signed flux averaged over regions where the structures 1, -1 and 2 evolve (Fig. \ref{fig:results:Bradflux_34_56}). These regions are those enclosed by a rectangle in green in Figs. \ref{fig:results:slit1_bmovement} and \ref{fig:results:slit2_bmovement}. To compute the averaged flux, we only selected the pixels three times above the Stokes-V noise (see Appendix \ref{sec:annex:Blos_confint} for more details).

First, we found small-scale flux emergence and rapid variations, a few minutes before the EUV emission enhancements measured with slit 1 ($E_1$ and $E_2$). A pair of negative and positive polarity structures, named $-1$ and 1 respectively, appeared co-temporally and next to each other (from 04:02:24 UT to 04:08:24 UT; Fig. \ref{fig:results:slit1_bmovement}). The $B_\mathrm{LOS}/\mu$ values of structures 1 and -1 reach \SI{33}{\gauss} and \SI{-44}{\gauss} at 04:05:24 UT, respectively. Their photospheric velocity vectors face opposite direction, which is consistent with the emergence of a small-scale magnetic bipole. Structure 1, with a positive polarity, moves toward the negative footpoint of loop bundle 1. The averaged flux shows the co-temporal increase of negative and positive fluxes up to \SI{8e16}{\maxwell} from 04:02 UT to 04:06 UT, followed by a decrease from 04:06 UT to 04:11 UT. (Fig. \ref{fig:results:Bradflux_34_56}a). This is consistent with an emerging bipole followed by small-scale field variations happening on both footpoints. As structure 1 with a positive polarity moves toward the negative footpoint of loop bundle 1 while its flux decreases, we consider the small-scale bipole rooted in structures 1 and -1 to be a potential candidate for a reconnection with loop bundle 1. This might result in EUV emission enhancements such as $E_1$ at 04:13 UT, and $E_2$ at 04:15 UT. See Sect. \ref{sec:discussion:thermal_plasma_flux} for a discussion on the subject.

Second, we observed two small-scale structures with positive polarity, called structures 2 and 3, which appeared \SI{5}{} and \SI{2}{\min} before the EUV emission enhancement measured with slits 2 and 3 (from 04:28 UT to 04:34 UT; Fig. \ref{fig:results:slit2_bmovement}), respectively. The $B_\mathrm{LOS}/\mu$ values of structures 2 and 3 reach respectivelly \SI{26}{\gauss} and \SI{30}{\gauss} at 04:31:24 UT. These structures are located below slit 3, and next to the larger-scale negative polarity footpoint of loop bundle 1. The magnetic structures 2 and 3 both have typical photospheric velocities of about \SI{3}{\kilo\meter\per\second} at maximum. The velocity vector of structure 2 is directed toward the negative footpoint of loop bundle 1. The signed flux of a region where structure 2 evolves shows a positive flux increases from 04:28 UT to 04:30 UT up to \SI{5e16}{\maxwell}, when the structure emerges (Fig. \ref{fig:results:Bradflux_34_56}b). Then, after 04:30 UT, we observe a decrease of the positive flux and rapid small-scale flux variation at the boundary of the footpoint. This variation happens at the time when $E_3$ and $E_4$ are measured with slit 2 (at 04:33 UT; Fig. \ref{sec:results:apparent_v}b). The negative flux of the footpoint summed over the region is not shown, as its variation was two order of magnitudes above the one of the positive flux ($\sim $ \SI{e19}{\maxwell}). We propose the hypothesis that the small-scale magnetic bipole rooted in magnetic structure 2 is a possible candidate for a reconnection with loop bundle 1, which might be at the origin of $E_3$ and $E_4$. See Sect. \ref{sec:discussion:thermal_plasma_flux} for a discussion on the subject.
 
\section{Discussion}
\label{sec:discussion}

We reported the detection of small-scale impulsive EUV emission enhancements along three bundles of network loops in a QS region. Combined observations of HRIEUV, PHI-HRT, and IRIS allowed us to derive the following properties for the EUV emission enhancements: they are common features in the network loops we have studied. Their intensity in HRIEUV reaches no more than 18\% above the background. We found four intensity peaks that seem to appear instantaneously along the slit (E$_1$, E$_2$, E$_4$, and E$_6$; Figs. \ref{fig:results:v_apparente_slit_124_world} and \ref{fig:results:v_apparente_slit_5_world}). We interpret them as being either truly instantaneous features or propagating features with PoS velocities above the limit resolvable by HRIEUV. This limit ranges from \SI{220}{\kilo\meter\per\second} for $E_2$ to \SI{510}{\kilo\meter\per\second} for $E_6$ (see Table. \ref{table:events}).

For two EUV emission enhancements, we also measured a subsonic ($\leq$ \SI[separate-uncertainty = true]{77(19)}{\kilo\meter\per\second}) intensity motion co-temporal with the fast component, which started from one of the footpoints of the loop. These slower PoS velocities are similar to those measured in HRIEUV by \cite{Mandal_2021}. These authors indeed measured HRIEUV brightenings propagating at a PoS velocity between \SI{25}{} and \SI{60}{\kilo\meter\per\second} along small-scale (3 to \SI{5}{\mega\meter}) loop-like structures. The slower apparent motions seen in intensity in our work are likely to be similar features in network loops. An important reminder is that a PoS motion of an HRIEUV intensity peak does not necessarily indicate mass motion. Nevertheless, we confirmed the presence of subsonic flows in the TR for one case, based on the measured Doppler shifts in the \ion{Si}{iv} \SI{1393.76}{\angstrom} line at the time of one EUV emission enhancement, and at the location of the right footpoint of loop bundle 1 (Sect. \ref{sec:results:doppler_chromosphere}). These shifts can be associated with upflows (up to \SI{32}{\kilo\meter\per\second}; Fig. \ref{fig:annex:iris_spec_event8_v2_specific_points}a) and downflows (up to \SI{18}{\kilo\meter\per\second}; Fig. \ref{fig:annex:iris_spec_event8_v2_specific_points}b) of plasma at lower TR temperatures ($\log{T} = 4.8$). We note that these upward velocities in the TR are consistent with previous observations of jets in IRIS \citep[\SI{20}{} to \SI{70}{\kilo\meter\per\second};][]{Gorman_2022}. Finally, with PHI-HRT, we noticed clear signatures of emergence and rapid flux variations of small-scale ($\approx$\SI{5e16}{\maxwell} to \SI{8e16}{\maxwell}) mixed-polarity magnetic structures near the footpoint of loop bundle 1 (Figs. \ref{fig:results:slit1_bmovement}, \ref{fig:results:slit2_bmovement}, and \ref{fig:results:Bradflux_34_56}). In one case, a magnetic bipole starts emerging \SI{3}{} to \SI{10}{\min} before the detection of EUV emission enhancements with HRIEUV, and the small-scale positive polarity footpoint decreases in flux while moving toward the larger-scale negative polarity footpoint of loop bundle 1 (Fig. \ref{fig:results:Bradflux_34_56}a).

Further discussion is divided into four main points. In Sect. \ref{sec:discussion:relation} we propose models explaining the combined results from HRIEUV and IRIS observations. In Sect. \ref{sec:discussion:thermal_plasma_flux} we review the mass and energy transfer models that can be responsible for the EUV emission enhancements as seen in HRIEUV and IRIS. In Sect. \ref{sec:discussion:photospheric_driving} we discuss the possible photospheric drivers of the EUV emission enhancements in HRIEUV. We also discuss the possible impact of the reconnection between the lower part of the loop bundle and small-scale magnetic bipoles. Finally, in Sect. \ref{sec:discussion:nextsteps} we propose possible next steps for this work.

\subsection{Connection between the results from HRIEUV and IRIS observations}
\label{sec:discussion:relation}

Together, HRIEUV and IRIS image plasma at a wide range of plasma temperatures, from the chromosphere to the corona. The plasma at these different temperatures might be unrelated to each other in terms of magnetic connection. However, in this section we provide evidence that suggests a connection between the intensity peak in the \ion{Si}{iv} \SI{1393.76}{\angstrom} line observed by IRIS, and the apparent motion seen in HRIEUV at 04:50 UT in event $E_7$ (Sect. \ref{sec:results:doppler_chromosphere}). First, the event occurs in the two bands at the same time and in the same region with a small spatial shift below \SI{1}{\mega\meter} (Fig. \ref{fig:results:iris_FOV_spec}). The latter can be explained by the difference of temperature, and thus height, of the emitting plasma. Stereoscopic effects can also be at play, due to the \SI{62.6}{\degree} pitch angle between Earth, Sun and Solar Orbiter. Another argument supporting the connection between IRIS and HRIEUV results is the location of the \ion{Si}{iv} intensity peak, at the right footpoint of loop bundle 1. As $E_7$ appears to originate from the footpoint, this is the location where one would expect the TR emission to be associated with $E_7$. Furthermore, the PoS velocity of $E_7$ (\SI[separate-uncertainty = true]{44}{\kilo\meter\per\second}; Fig. \ref{fig:results:v_apparente_slit_124_world}c) is in the same range of the LOS velocity upflows (\SI{32}{\kilo\meter\per\second}; Fig. \ref{fig:annex:iris_spec_event8_v2_specific_points}) and downflows values (up to \SI{18}{\kilo\meter\per\second}) measured with the \ion{Si}{iv} line. As the intensity peak motion of $E_7$ is directed from the right footpoint toward the left footpoint, it is more likely to be associated with the upflows than with the downflows. The downflows have no equivalent in apparent motion seen in HRIEUV. This is to be expected, because the downflows directed toward the chromosphere can be associated with higher density and cooler plasma compared to the upflows \citep[][]{Chen_2022}. Thus, the cooler downflows would only be observed in the \ion{Si}{iv} line ($\log{T} = 4.8$), and not in HRIEUV (from $\log{T} = 5.4$ to $6.0$). As such, the $E_7$ and the \ion{Si}{iv} Doppler shifts are good candidates to be the higher TR/coronal and lower TR signatures of plasma flows originating from lower part of loop bundle 1, respectively.

\subsection{Thermal flux and plasma flows, caused by an impulsive energy deposition}
\label{sec:discussion:thermal_plasma_flux}

In this section we combine the multi-instrumental results discussed in the previous section to present model ideas that can explain the physical origins of the EUV emission enhancements along the network loops, in term of plasma flows and thermal effect. In this work, we measured four events with unresolved PoS velocities. The latter are either truly instantaneous features or are propagating features with PoS velocities above the limit due to the HRIEUV cadence (from \SI{220}{\kilo\meter\per\second} to \SI{520}{\kilo\meter\per\second}). In all cases, these are not features propagating at sound speed \citep[$c_\mathrm{s} = $ \SI{152}{\kilo\meter\per\second} at \SI{1}{\mega\kelvin};][]{Priest_1982}. Thus, unlike PDs for instance, their PoS velocity is not consistent with slow-magneto-acoustic waves \citep[][]{Kiddie_2012,Prasad_2014}. Therefore, we focus on four possibilities to explain EUV emission enhancements: impulsive heating located at the basis of the loop similar to spicules; nanoflare heating in the coronal part of the loop; syphon flows; and TI-TNE cycles. This list is non-exhaustive and only applies to the specific EUV emission enhancements studied in Sects. \ref{sec:results:apparent_v} to \ref{sec:results:magnetic}.

One key feature on the PoS velocities for two of the EUV emission enhancements studied in this work is the co-temporal appearance of intensity peak motions with an unresolved ($v \geq$ \SI{220}{\kilo\meter\per\second}) and a resolved PoS velocity ($v\leq$ \SI{77}{\kilo\meter\per\second}) along the loop bundle (Figs. \ref{fig:results:v_apparente_slit_124_world}c and \ref{fig:results:v_apparente_slit_5_world}). Fast apparent motions seen in UV-EUV imagers or spectro-imagers have been previously observed for different types of atmospheric events, including type-II spicules \citep[\SI{80}{} to \SI{350}{\kilo\meter\per\second};][]{Tian_2014,Narang_2016} or propagating disturbances along coronal plumes \citep[\SI{82}{} to \SI{160}{\kilo\meter\per\second};][]{Huang2025}. In particular, \cite{de_pontieu_2017} used 3D magnetohydrodynamics (MHD) simulations to understand the origin of the discrepancy in type-II spicules between the low LOS Doppler velocity measured in chromospheric lines \citep[\SI{20}{} to \SI{50}{\kilo\meter\per\second};][]{Sekse_2012,Sekse_2013} and the fast PoS velocities measured by SJI at lower TR temperature (up to \SI{350}{\kilo\meter\per\second}). These results showed that the high PoS velocities in SJI images can be explained by the propagation of a heating front at Alfv\'en velocities, in contrast to the mass motions measured with the Doppler shifts. The authors concluded that this heating front existed because ``electric current propagates along the magnetic field at Alfv\`enic speed, driven by tension and/or transverse waves.''

The first model scenario we propose for the events we see relies on the similarities between the EUV emission enhancements and spicules: a fast and a slow component measured with UV-EUV imagers; LOS velocities measured with the Doppler shift of TR lines close to the PoS velocity of the slow component; and signatures of magnetic cancellation at the footpoints \citep[][]{Samanta_2019}. As such, we propose a hypothesis similar to the model of spicules to explain our results. First, the network loop bundle reconnects at its basis through internal braiding or the reconnection with small-scale magnetic structures. This leads to the propagation of a heating front, seen as a fast apparent motion in HRIEUV ($E_1$, $E_2$, $E_4$, and $E_6$). The propagation of this heating front can be driven by a similar mechanism as the one described in \cite{de_pontieu_2017} or by an alternative one, such as thermal conduction or a MHD shock. The increased pressure at the basis of the loop following the energy deposition drives slower (subsonic) plasma flows, seen as PoS motions ($E_3$, $E_5$, and $E_7$) in HRIEUV (Figs. \ref{fig:results:v_apparente_slit_124_world}, \ref{fig:annex:v_apparente_slit_3_world}, and \ref{fig:results:v_apparente_slit_5_world}). In particular, the bi-directional jet seen in HRIEUV ($E_5$; Fig. \ref{fig:annex:v_apparente_slit_3_world}) and in the Doppler shift of the \ion{Si}{iv} lines (Fig. \ref{fig:results:iris_spec_event8}) are strong evidence of a localized energy deposition in the lower part of the loop that might be compared to explosive events \citep[e.g.,][]{Innes_1997}.

For the second model scenario we propose that impulsive energy deposition (e.g., nanoflares) heat up the upper part of a loop strand already filled up with plasma. During the heating or the cooling time, the HRIEUV intensity increases all along the loop almost co-temporally, when the temperature reaches the peak of the HRIEUV response function (\SI{1}{\mega\kelvin}). This would explain the fast intensity motions ($E_1$, $E_2$, $E_4$, and $E_6$). Then, the chromosphere or the lower TR gets heated from the corona through conduction \citep[][]{dolliou_2025} or non thermal particles \citep[][]{Testa_2014} leading to bi-directional plasma flows along the field lines. Similar to the first model scenario, this would explain the slower intensity motions ($E_3$, $E_5$, and $E_7$) and the Doppler velocities measured in the \ion{Si}{iv} lines with IRIS.

The third model scenario is based on siphon flows. They occur regularly along the loop when the spatial distribution of the heating between the two footpoints is asymmetrical \citep[e.g.,][]{Klimchuk_Luna_2019}. The pressure gradient results in plasma flows propagating from one footpoint to the next. \cite{Bethge_2012} observed these siphon flows on a loop bundle with similar properties ($\approx$ \SI{21}{\mega\meter} in length) as the ones studied in this work. The authors measured a Doppler velocity of about \SI{40}{\kilo\meter\per\second} in the chromospheric \ion{He}{i} \SI{1083.03}{\nano\meter} line. This value is close to the velocities of the slower components we measured with the apparent motions in HRIEUV ($44 \pm 9$ \SI{}{\kilo\meter\per\second}) and with the Doppler shift of the \ion{Si}{iv} line (up to \SI{32}{\kilo\meter\per\second}). However, in their specific magnetic configuration, they estimated the theoretical siphon flow velocity to reach \SI{140}{\kilo\meter\per\second} at maximum. Assuming our loop magnetic configuration to be comparable, this velocity value is too low to explain the events appearing instantaneously or propagating with high PoS velocities ($\geq$ \SI{220}{\kilo\meter\per\second}) we measured in HRIEUV.

Finally, TI-TNE cycles can also produce periodic intensity increases in EUV imagers \citep[][]{Froment_2017}. We note that TI-TNE cycles refer to the plasma response to either deviation from thermal equilibrium \citep[TI;][]{Klimchuk_2019} or constraints on the spatial distribution of the heating along the loop \citep[][]{Klimchuk_Luna_2019}. TI-TNE cycles models are at least partly consistent with the first two model scenarios involving an impulsive energy release in the upper or the lower part of the loop. Impulsive heating can indeed produce TI by shifting the loop out of equilibrium, and regular impulsive energy deposition near the footpoint can also contribute to the development of a TNE cycle.

\subsection{Photospheric motions of loop footpoints or emergence of small-scale magnetic bipole as drivers of EUV emission enhancements}
\label{sec:discussion:photospheric_driving}

In this section we aim to refine the discussion on the physical origin of EUV emission enhancements. Assuming they originate from magnetic reconnection, we propose two (non-exclusive) models : reconnection at the loop bundle basis with small-scale bipoles ; and "internal" braiding of field lines within the loop bundle.  

Our results from PHI-HRT confirm the emergence and rapid flux variation of small-scale opposite-polarity magnetic field near the negative polarity footpoint of loop bundle 1 during the time we observed it (Sect. \ref{sec:results:magnetic}). As en example, Fig. \ref{fig:results:slit8_bmovement} shows a large number of small-scale magnetic structures with a positive polarity above the $4\sigma$ noise level. Small-scale flux emergence in the intra-network and in the QS has already reported in other works \citep[][]{gosic_2022,Chitta_2023}. In particular, \cite{Chitta_2017} observed similar small-scale mixed-polarity structures at the footpoint of AR loops. We found two cases of small-scale flux emergence and rapid variation. This is consistent with small magnetic structures emerging and then reconnecting with loop bundle 1 (Sect. \ref{sec:discussion:relation}). These small-scale magnetic structures could then reconnect with the lower part of the loop bundle. The separation timescale of a few minutes between the EUV emission enhancements is also of the order of those of photospheric granulation for the flux emergence \citep[\SI{5}{} to \SI{10}{\min};][]{Hirzberger_1999} and of p modes \citep[\SI{5}{\min};][]{Leighton_1962}, so that they could be related to the reconfigurations of the field lines.

Another (non-exclusive) possible physical origin for these EUV emission enhancement is the "internal" braiding of magnetic strands within the loop bundles, driven by photospheric motions. In that regards, \cite{Breu_2022} simulated the thermal and mass transfers in a coronal loop (of \SI{50}{\mega\meter} length). The results showed that the photospheric motion on the loop footpoints provided enough energy to sustain the loop. The total (viscous and Ohmic) heating peaked impulsively, with a separation timescale of a few minutes. The electron density $n_\mathrm{e}$ and temperature $T_\mathrm{e}$ profiles along the loop shared the bursty behavior of the heating, driven by the regular relaxation of braided loop strands. When we extrapolate these results to our specific cases (for instance, loop bundle 1), variation with time in the $n_\mathrm{e}$ and $T_\mathrm{e}$ profiles are likely to have an impact to the HRIEUV light curve. The differential emission measure (DEM) of CBPs has been measured with EUV spectroscopy and was found to peak at temperatures between $\log{T} = 6.1$ to $6.3$ \citep[][]{Golub_1974,Alexander_2011,Milanovic_2025}. As the response function of HRIEUV has its maximum at $\log{T} = 6.0$, we can expect to detect the emission from plasma all along the loop. Density fluctuations and thermal effects will then have a significant impact on the HRIEUV intensity modulations.

\subsection{Next steps} 
\label{sec:discussion:nextsteps}

We propose two possible follow-ups to this study. First, increasing the selected pool of four EUV emission enhancements to a statistically significant number. This is possible, because they are likely to be present at all times along network loop bundles. Statistical results at higher HRIEUV cadence could be useful to evaluate whether they originate from a single or multiple physical origins. Another important follow-up would be to better evaluate the plasma properties (temperature, density) associated with the plasma emitting these EUV emission enhancements. HRIEUV is sensitive to the emission of plasma over a rather broad temperature range, from $\log{T} [{\rm{K}}]= 5.4$ to $6.0$. Constraining the plasma temperature and density \citep[e.g.,][]{Dolliou_2023,Dolliou_2024,Huang_2023} is necessary to estimate the contribution of the impulsive heating to the loop energy budget. This calls for high spatial and temporal resolution spectroscopy achieved simultaneously over a wide temperature range, a design characteristic of the Solar-C/EUV High-throughput Spectroscopic Telescope \citep[EUVST;][]{Shimizu_2019} currently in its implementation phase. These studies will become possible thanks to its high spatial (\SI{0.4}{\arcsec}) resolution, and its wavelength coverage, which includes lines emitted by plasma from chromospheric ($\log{T} = 4.3$) to flare temperatures ($\log{T} = 7.0$)

\section{Conclusions}
\label{sec:conclusions}

We reported small-scale emission enhancements in the EUV in small network loops in a multi-instrument study giving access to the magnetic field in the photosphere, the emission and line spectra in the transition region, and the emission from the corona. The key observational result is the detection of a brightening that seems to appear almost instantaneously (within the limited time resolution of the observations) along the slit, followed by a brightening with a resolved PoS motion. The multi-instrumental approach allowed us to detect associated Doppler shifts in the \ion{SI}{iv} line and cases of small-scale flux emergence and disappearance near the footpoint of the loop bundle.

The properties of the observed brightenings suggest that the observations can be understood through several model scenarios.
One consistent scenario is similar to that of spicules in that a heating event low in the loop causes a fast heating front, followed by a slow motion in response to the pressure enhancement.
In another scenario, the heating would appear higher up in the loop and almost instantaneously cause a brightening of the plasma. Similar to chromospheric evaporation, the downward heat conduction would then lead to a slow filling of the loop.
Further scenarios can be envisioned, but seem less applicable at this point. 
The ultimate source of the EUV brightenings might be leakage of p modes, granulation motions, shuffling of lines through photospheric motions, and the interaction of emergence and cancellation of small-scale magnetic concentrations.

\begin{acknowledgements}
The authors thank the anonymous referee for the suggestions that greatly helped to improve the clarity of the manuscript. The authors would like to thank G.~Valori, J.~Sinjan and D.~Przybylski for the very fruitful discussions and L.~R.~Bellot~Rubio for coordinating the SOOP. The work of A. Dolliou is funded by the Federal Ministry for Economic Affairs and Climate Action (BMWK) through the German Space Agency at DLR based on a decision of the German Bundestag (Funding code: 50OU2101, 50OU2201). L.P.C. gratefully acknowledges funding by the European Union (ERC, ORIGIN, 101039844). Y. Chen acknowledges funding provided by the Alexander von Humboldt Foundation. Solar Orbiter is a space mission of international collaboration between ESA and NASA, operated by ESA. The EUI instrument was built by CSL, IAS, MPS, MSSL/UCL, PMOD/WRC, ROB, LCF/IO with funding from the Belgian Federal Science Policy Office (BELSPO/PRODEX PEA 4000112292 and 4000134088); the Centre National d’Etudes Spatiales (CNES); the UK Space Agency (UKSA); the Bundesministerium für Wirtschaft und Energie (BMWi) through the Deutsches Zentrum für Luft- und Raumfahrt (DLR); and the Swiss Space Office (SSO). We are grateful to the ESA SOC and MOC teams for their support. The German contribution to SO/PHI is funded by the BMWi through DLR and by MPG central funds. The Spanish contribution is funded by AEI/MCIN/10.13039/501100011033/ and European Union “NextGenerationEU”/PRTR” (RTI2018-096886-C5,  PID2021-125325OB-C5,  PCI2022-135009-2, PCI2022-135029-2) and ERDF “A way of making Europe”; “Center of Excellence Severo Ochoa” awards to IAA-CSIC (SEV-2017-0709, CEX2021-001131-S). The French contribution is funded by CNES. IRIS is a NASA small explorer mission developed and operated
by LMSAL with mission operations executed at NASA Ames Research center and
major contributions to downlink communications funded by ESA and the
Norwegian Space Centre. This research used version 7.0.1 \citep{stuart_j_mumford_2025_16638197} of the SunPy open source software package \citep{sunpy_community2020}. 
\end{acknowledgements}

\bibliographystyle{aa}
\bibliography{Biblio.bib}

\begin{appendix}

\section{Method for computing the PoS velocities in HRIEUV time-distance maps}
\label{sec:annex:measurement_apparent_v}

\begin{figure*}
    \includegraphics{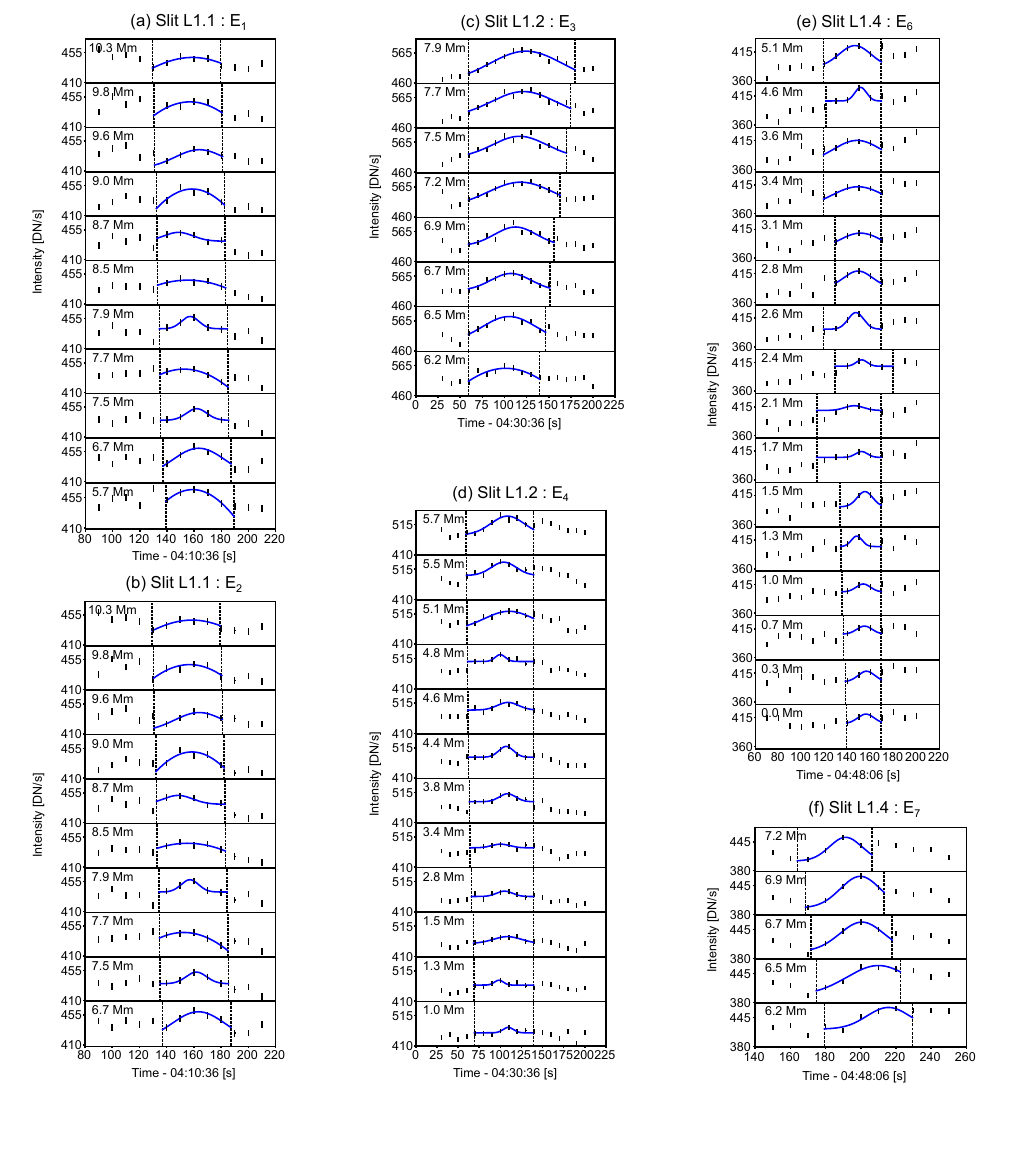}
    \caption{Intensity profiles at specific locations along slits L1.1 to L1.4 for six events (from $E_1$ to $E_7$, with the exception of $E_5$) shown in Figs. \ref{fig:results:v_apparente_slit_124_world} and \ref{fig:results:v_apparente_slit_5_world}. The blue curve shows the fitted Gaussian within the time interval (dotted vertical line). The locations shown are those with an uncertainties on the central time below 15 seconds. The locations correspond to the red crosses displayed in Figs. \ref{fig:results:v_apparente_slit_124_world} and \ref{fig:results:v_apparente_slit_5_world}.}
    \label{fig:annex:intensity_profiles_slit_1245}
\end{figure*}

\begin{figure}
    \centering
    \includegraphics{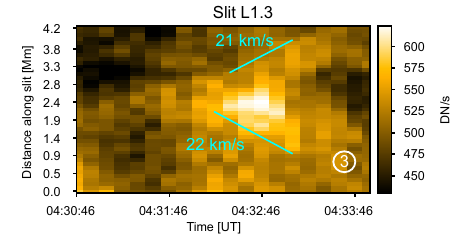}
    \caption{Bi-directional intensity motion in HRIEUV ($E_5$), measured with slit L1.3 (Fig. \ref{fig:results:all_slits_fov_oneimage}c), and their estimated velocities (in cyan). }
    \label{fig:annex:v_apparente_slit_3_world}
\end{figure}

\renewcommand{\arraystretch}{1.5} 
\begin{table}
\caption{Events reported in this work.}    
\flushleft          
\begin{tabular}{l c c} 
\hline
\hline
Event & $v_\mathrm{PoS}$ [\SI{}{\kilo\meter\per\second}] & $v_\mathrm{res}$ [\SI{}{\kilo\meter\per\second}]\\
\hline
$E_1$ & NR &  463 \\
$E_2$ & NR &  226 \\
$E_3$ & $ 77 \pm 19 $ &  175 \\
$E_4$ & NR &  463 \\
$E_5$ & 21 and 22 & NA \\
$E_6$ & NR &  516 \\
$E_7$ & $ 49 \pm 9 $ & 103 \\
\end{tabular}      
\tablefoot{Velocities estimated for the events studied in this work. For each event (see Figs. \ref{fig:results:v_apparente_slit_124_world} and \ref{fig:results:v_apparente_slit_5_world}), the table provides its name, its computed PoS velocity $v_\mathrm{PoS}$ and the limit velocity that can be resolved given the HRIEUV cadence $v_\mathrm{res}$. Events with a PoS velocity that could not be temporally resolved are marked as "NR" (Not Resolved). The PoS velocity of $E_5$ was the only one to be estimated visually, and not with the algorithm described in Appendix \ref{sec:annex:measurement_apparent_v}.}
\label{table:events}  
\end{table}

In this section, we describe how PoS velocities of the HRIEUV intensity peaks associated with the four EUV emission enhancements are computed on time-distance maps (Figs. \ref{fig:results:v_apparente_slit_124_world} and \ref{fig:results:v_apparente_slit_5_world}). As indicated in Sect. \ref{sec:results:apparent_v}, we define as events the intensity peaks in time-distance maps that either propagate or seem to appear instantaneously along the slit, under the limitations of the HRIEUV cadence of \SI{10}{\second}. The following method was applied from events $E_1$ to $E_7$, with the exception of $E_5$. The latter is discussed at the end of this section. 

Because their intensity peak is low compared to the background, the events can be hard to distinguish on time-distance maps (e.g., $E_1$ and $E_2$ in Fig. \ref{fig:results:v_apparente_slit_124_world}). Thus, we identified their intensity peaks directly in the HRIEUV light curves at distinct locations along the slit (Fig. \ref{fig:annex:intensity_profiles_slit_1245}). The uncertainties of the HRIEUV light curves were computed from the readout (2 DN) and the photon noise \citep[more details in][]{euidatarelease6}. At each location, we created time intervals that included the intensity peaks associated with the events (dotted vertical black lines in Fig. \ref{fig:annex:intensity_profiles_slit_1245}). We fit the light curves within the time interval with a Gaussian function (blue curves in Fig. \ref{fig:annex:intensity_profiles_slit_1245}). The method implies that the shape of the intensity peak is close to a Gaussian function. This assumption is reasonable in most cases, as the HRIEUV point spread function (PSF) tends to smooth the intensity peaks toward such a shape. From the fit results, we computed the central time of the peak $t_\mathrm{peak}$ and its uncertainty. We excluded the locations for which the fitting was unable to accurately constrain the central times. These locations are those with uncertainties on the central time larger than 15 seconds.

For each event, we performed a linear fit to the function $t_\mathrm{peak}(l)$, with $l$ being the location along the slit. We also included the uncertainty of $t_\mathrm{peak}$ to the fitting. The velocities and their uncertainties are derived from the inverse of the slope and the propagation of the slope error. The event E$_4$ (Fig. \ref{fig:results:v_apparente_slit_124_world}a) is a special case, as the value on the slope of the linear fit is close to zero. The uncertainty on the slope switches the velocity values from negative to positive infinity. In that case, we set a lower limit  $v_\mathrm{low}$ defined as the minimum among the absolute values of the velocity interval endpoints. We note that it is not possible to determine the direction of propagation along the slit for E$_4$. 

We also define a lower limit $v_\mathrm{res} = D/\Delta t$ due to the temporal resolution of HRIEUV ($\Delta t=$\SI{10}{\second}). The distance $D$ is the one over which the event velocity is measured. When the lower limit of the velocity due to the uncertainty of the  least-squares fit or $v_\mathrm{low}$ was above $v_\mathrm{res}$, then we considered the event velocity to be unresolved. As a remark, all lower limits shown in Figs. \ref{fig:results:v_apparente_slit_124_world} and \ref{fig:results:v_apparente_slit_5_world} are due to the temporal resolution of HRIEUV. 

The event $E_5$ measured along slit 3 (Fig. \ref{fig:annex:v_apparente_slit_3_world}) corresponds to two intensity peaks moving in opposite directions near the right footpoint of loop bundle 1, at the time of the EUV emission enhancement measured with slit 2. In this case, the complexity of the light curves on specific positions prevented us from fitting Gaussian functions. Instead, we visually measured the PoS velocities of \SI{21}{} and \SI{22}{\kilo\meter\per\second}. The results for all events are shown in Table. \ref{table:events}.

Finally, we measured the ratio of intensity peak to the background value for the events $E_4$, $E_6$ and $E_7$, where the background intensity could be accurately estimated. We computed the ratio of the fitted Gaussian peak to the background level at all locations where we measure the events. The results showed that the intensity peaks of the events  $E_4$, $E_6$, and $E_7$ reach values no higher than 14\%, 12\%, and 18\%, respectively, above the background level.

\section{Confidence levels in $B_\mathrm{los}$ images measured by PHI-HRT}
\label{sec:annex:Blos_confint}

 The magnetic structures investigated in Sect. \ref{sec:results:magnetic} can reach small spatial, temporal and magnetic scales (down to  $\Delta x =$ \SI{0.2}{\mega\meter}, $\Delta t =$ \SI{3}{\min} and $B_\mathrm{los}/\mu =$ \SI{21}{\gauss}). As these scales are close to the instrumental limitations of PHI-HRT, we compute confidence levels to ensure that the magnetic structures cannot be explained by noise alone. We chose to evaluate the noise on the Stokes-V vector, as $B_\mathrm{los}$ mainly depends on the circular polarization, and thus Stokes-V. The confidence levels (pink contours) on the $B_\mathrm{los}/\mu$ images represent the pixels above the $4\sigma$ Stokes-V noise threshold:
\begin{equation}
\label{eq:results:Blos_conf}
    \mid V \mid_\mathrm{max} > 4\sigma_\mathrm{v,cont} + \mu_\mathrm{v,cont}
\end{equation}
$\mid V \mid_\mathrm{max}$ is the maximum absolute value of Stokes-V over the five non-continuum wavelengths for the pixel. $\mu_\mathrm{v,cont}$ and $\sigma_\mathrm{v,cont}$ are the average and the standard deviation, respectively, over all pixels of the Stokes-V values at the near-continuum wavelength. As Stokes-V is supposed to be dominated by noise in the continuum, the condition \ref{eq:results:Blos_conf} ensures that the $B_\mathrm{los}$ values of all pixels within the confidence levels have a likelihood lower than 0.007\% to be explained by noise alone. Furthermore, the lifetime of every magnetic structure of interest is longer than three images, which decreases the probability even more that of them is due to noise. The confidence limits appear as the $4\sigma$ pink contours in Figs. \ref{fig:results:slit1_bmovement}, \ref{fig:results:slit2_bmovement}, and \ref{fig:results:slit8_bmovement}. In Figs. \ref{fig:results:Bradflux_34_56} and \ref{fig:annex:Bradflux_8}, only pixels above the $3\sigma$ Stokes-V noise level are selected to compute the signed fluxes. The noise threshold has been decreased from $4\sigma$ to $3\sigma$ to include the variations at lower values of $B_\mathrm{los}$, while keeping a reasonable confidence level.

\section{Small-scale opposite polarity flux emergence during the EUV emission enhancement along slit L1.4}
\label{sec:annex:magnetic_slit5}
\begin{figure*}
    \includegraphics{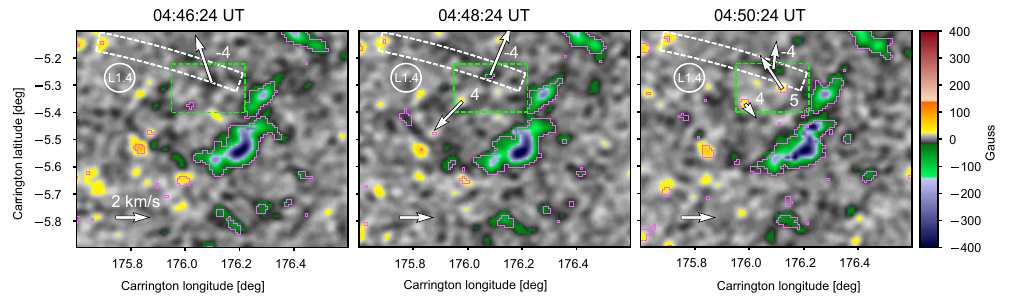}
    \caption{Same as Fig. \ref{fig:results:slit1_bmovement} but for slit L1.4 (Fig. \ref{fig:results:all_slits_fov_oneimage}d).}
    \label{fig:results:slit8_bmovement}
\end{figure*}

\begin{figure}
    \centering
    \includegraphics{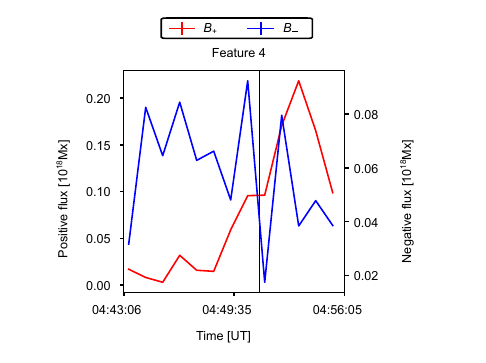}
    \caption{Same as Fig. \ref{fig:results:Bradflux_34_56} but for the signed flux averaged over the green rectangle of Fig. \ref{fig:results:slit8_bmovement}.}
    \label{fig:annex:Bradflux_8}
\end{figure}

Figure \ref{fig:results:slit8_bmovement} shows $B_\mathrm{los}/\mu$ images at three different times before the EUV emission enhancements measured with slit L1.4 (04:50 UT; Fig. \ref{fig:results:v_apparente_slit_5_world}). We observe small-scale flux emergence all around the negative footpoint of loop bundle 1. In particular, the positive (4) and negative-polarity (-4) structures of interest seem to emerge as a bipole. Their flux increase at the same time, and their velocity is oriented in the opposite direction. However, we were unable to measure any specific flux emergence and rapid variation that could be related to the EUV emission enhancement measured with slit L1.4 (Fig. \ref{fig:results:v_apparente_slit_5_world}). Figure \ref{fig:annex:Bradflux_8} displays the signed flux averaged over the region enclosed by a rectangle in green, which is at the location of slit L1.4. we see an increase in the positive flux, while the variation in the negative flux is unclear.

\section{Additional figures}
\label{sec:annex:additional_figures}

\begin{figure*}
    \includegraphics{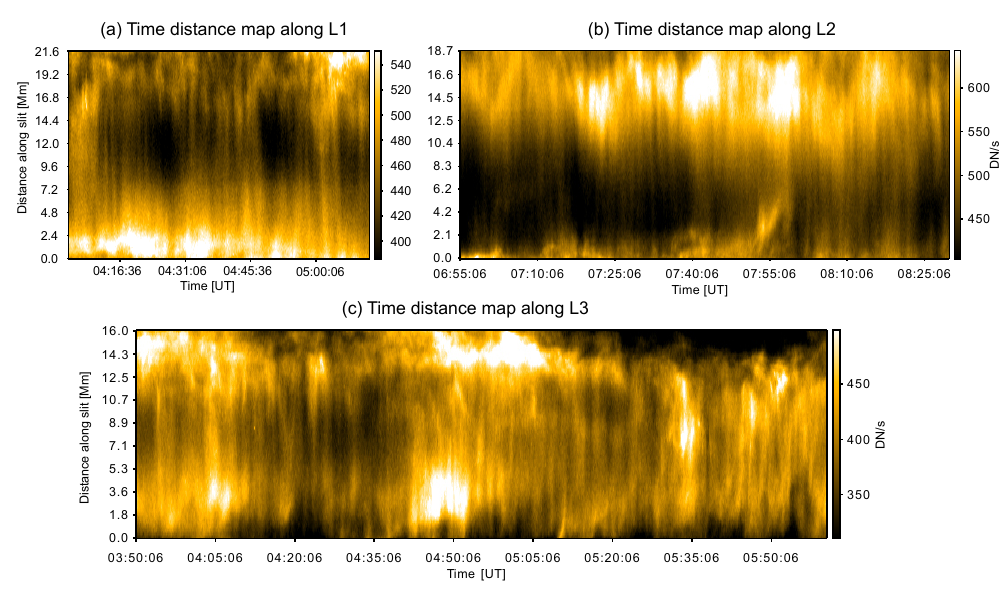}
    \caption{Time-distance maps of the loop bundles 1 (a), 2 (b), and 3 (c) computed along slits L1, L2, and L3, respectively, displayed in Fig. \ref{fig:results:loop1_2_3_fov}. No detrending has been applied to obtain these figures.}
    \label{fig:results:loop1_2_3_stackplots}
\end{figure*}

\begin{figure*}
    \centering
    \includegraphics{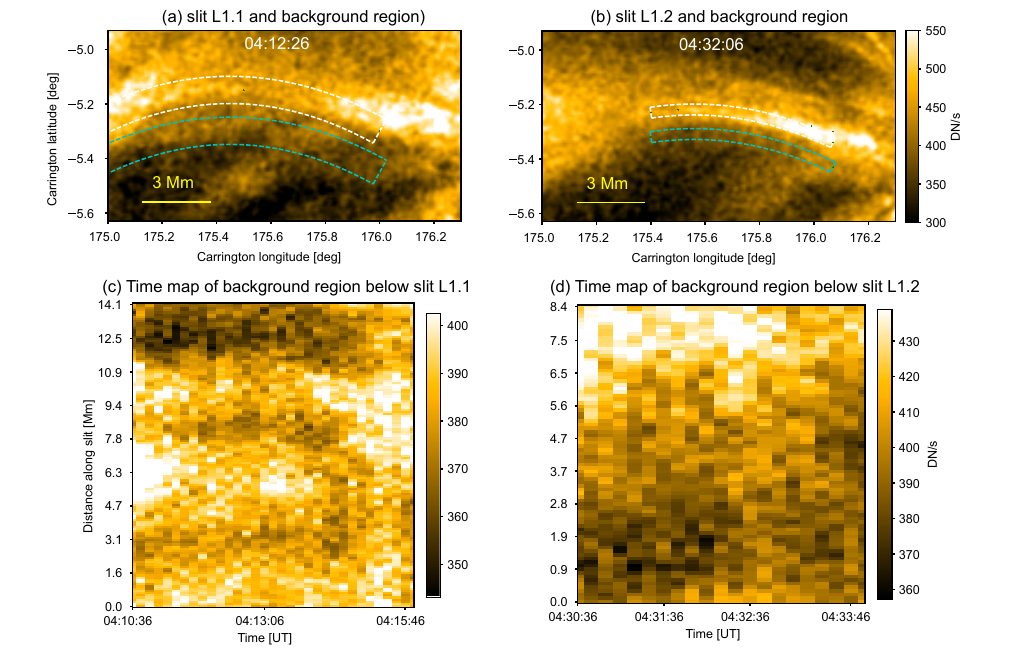}
    \caption{Estimation of the background emission below loop bundle 1 for the EUV emission enhancements measured with slits L1.1 and L1.2 (Figs. \ref{fig:results:v_apparente_slit_124_world}a and \ref{fig:results:v_apparente_slit_124_world}b). HRIEUV images of loop bundle 1 show the location of slit L1.1 (a) and L1.2 (c). Background slits, indicated by cyan dotted lines, are obtained by shifting the original slits (dashed white lines) in the Carrington latitude. The time-distance maps along the background slits are shown in panels (b) and (d) for slits L1.1 and L1.2, respectively. The EUV emission enhancements associated with the two slits are not detected in the background emission. }
    \label{fig:annex:background_slit_34}
\end{figure*}

We here propose a short description of the additional figures that are referred in the main text. In Fig. \ref{fig:results:loop1_2_3_stackplots} we show the time-distance maps along L1, L2 and L3 (Fig. \ref{fig:results:loop1_2_3_fov}) with the original data without detrending. About two dozens of EUV emission enhancements per hour are detected along the three loop bundles. The figure is referred to in Sect. \ref{sec:results:impulsive_qpp}. In Fig. \ref{fig:annex:background_slit_34} we verify that the EUV emission enhancements measured with slits L1.1 and 1.2 are not due to background emission below or above loop bundle 1 along the line of sight. Slits L1.1 and L1.2 are displayed on HRIEUV images in Figs. \ref{fig:results:all_slits_fov_oneimage}a and \ref{fig:results:all_slits_fov_oneimage}b, respectively. The background emission for slits L1.1 and L1.2 is estimated by creating spatially shifted slits outside of loop bundle 1 (Fig. \ref{fig:annex:background_slit_34}). We see that the resulting time-distance maps (Figs. \ref{fig:annex:background_slit_34}b and \ref{fig:annex:background_slit_34}d) do not display the events shown in Figs. \ref{fig:results:v_apparente_slit_124_world}a and \ref{fig:results:v_apparente_slit_124_world}b. The figure is referred to in Sect. \ref{sec:results:apparent_v}.

\end{appendix}
\end{document}